\documentclass[fleqn,usenatbib]{mnras}

\defcitealias{4FGL}{4FGL}
\defcitealias{4FGL-DR3}{4FGL DR3}
\defcitealias{YMW16}{YMW16}
\usepackage{newtxtext,newtxmath}

\usepackage[T1]{fontenc}
\usepackage{upgreek}

\DeclareRobustCommand{\VAN}[3]{#2}
\let\VANthebibliography\thebibliography
\def\thebibliography{\DeclareRobustCommand{\VAN}[3]{##3}\VANthebibliography}

\usepackage{graphicx}	% Including figure files

\newcommand{\oneF}{\mathcal{F}}

\newcommand{\fgw}{803.5}
\newcommand{\fdotgw}{-1.14 \times 10^{-15}}

\newcommand{\pvalue}{29}%percent
\newcommand{\hULval}{1.25 \times 10^{-26}}
\newcommand{\hULpercentUncertainty}{8} %percent

\newcommand{\ulratio}{20}

\newcommand{\epUL}{2.45\times 10^{-8}}

\title[TRAPUM's Survey of \textit{Fermi}-LAT sources]{The TRAPUM L-band survey for pulsars in \textit{Fermi}-LAT gamma-ray sources}

\author[C. J. Clark, et al. ]{
C. J. Clark$^{1,2,3}$\thanks{E-mail: colin.clark@aei.mpg.de}, R. P. Breton$^{3}$, E. D. Barr$^{4}$,
M. Burgay$^{5}$, T. Thongmeearkom$^{3}$, L. Nieder$^{1,2}$, \newauthor S. Buchner$^{6}$,
B. Stappers$^{3}$, M. Kramer$^{4,3}$, W. Becker$^{7,4}$, M. Mayer$^{7}$, A. Phosrisom$^{3}$,
A. Ashok$^{1,2}$,\newauthor M. C. Bezuidenhout$^{3}$, F. Calore$^{8}$, I. Cognard$^{9,10}$,
P. C. C. Freire$^{4}$, M. Geyer$^{6}$, J.-M. Grie{\ss}meier$^{9,10}$, \newauthor
R. Karuppusamy$^{4}$, L. Levin$^{3}$, P. V. Padmanabh$^{4,1,2}$, A. Possenti$^{5}$,
S. Ransom$^{11}$, M. Serylak$^{12,13}$, \newauthor V.~Venkatraman~Krishnan$^{4}$,
L. Vleeschower$^{3}$, J. Behrend$^{4}$, D. J. Champion$^{4}$, W. Chen$^{4}$,
D. Horn$^{6}$, \newauthor E. F. Keane$^{14}$, L. K\"{u}nkel$^{15}$, Y. Men$^{4}$,
A. Ridolfi$^{5,4}$, V. S. Dhillon$^{16,17}$, T. R. Marsh$^{18}$, M. A. Papa$^{1,2}$\\
$^{1}$ Max Planck Institute for Gravitational Physics (Albert Einstein Institute), D-30167 Hannover, Germany\\
$^{2}$ Leibniz Universit\"{a}t Hannover, D-30167 Hannover, Germany\\
$^{3}$ Jodrell Bank Centre for Astrophysics, Department of Physics and Astronomy, The University of Manchester, Manchester M13 9PL,
UK\\
$^{4}$ Max-Planck-Institut f\"{u}r Radioastronomie, Auf dem H\"{u}gel 69, D-53121 Bonn, Germany\\
$^{5}$ INAF -- Osservatorio Astronomico di Cagliari, Via della Scienza 5, I-09047 Selargius (CA), Italy\\
$^{6}$ South African Radio Astronomy Observatory, 2 Fir Street, Black River Park, Observatory 7925, South Africa\\
$^{7}$ Max-Planck-Institut f\"{u}r Extraterrestrische Physik, Giessenbachstra\ss{}e, 85748 Garching, Germany\\
$^{8}$ Laboratoire d'Annecy-le-Vieux de Physique Th\'{e}orique (LAPTh), CNRS, USMB, F-74940 Annecy, France\\
$^{9}$ LPC2E - Universit\'{e} d'Orl\'{e}ans / CNRS, 45071 Orl\'{e}ans cedex 2, France\\
$^{10}$Observatoire Radioastronomique de Nan\c{c}ay (ORN), Observatoire de Paris, Universit\'{e} PSL, Univ Orl\'{e}ans, CNRS, 18330
Nan\c{c}ay, France\\
$^{11}$ National Radio Astronomy Observatory, 520 Edgemont Rd., Charlottesville, VA 22903, USA\\
$^{12}$ SKA Observatory, Jodrell Bank, Lower Withington, Macclesfield, SK11 9FT, United Kingdom\\
$^{13}$ Department of Physics and Astronomy, University of the Western Cape, Bellville, Cape Town, 7535, South Africa\\
$^{14}$ School of Physics, Trinity College Dublin, College Green, Dublin 2, Ireland\\
$^{15}$ Fakult\"{a}t f\"{u}r Physik, Universit\"{a}t Bielefeld, Postfach 100131, D-33501 Bielefeld, Germany\\
$^{16}$ Department of Physics and Astronomy, University of Sheffield, Sheffield S3 7RH, UK\\
$^{17}$ Instituto de Astrof\'{i}sica de Canarias, E-38205 La Laguna, Tenerife, Spain\\
$^{18}$ Department of Physics, University of Warwick, Coventry CV4 7AL, UK\\ }

\date{Accepted 2022 December 13. Received 2022 December 13; in original form 2022 October 14}

\pubyear{2022}

\begin{document}
\label{firstpage}
\pagerange{\pageref{firstpage}--\pageref{lastpage}}
\maketitle

\begin{abstract}
  More than 100 millisecond pulsars (MSPs) have been discovered in radio observations of gamma-ray sources detected by the \textit{Fermi} Large Area Telescope (LAT), but hundreds of pulsar-like sources remain unidentified. Here we present the first results from the targeted survey of \textit{Fermi}-LAT sources being performed by the Transients and Pulsars with MeerKAT (TRAPUM) Large Survey Project. We observed 79 sources identified as possible gamma-ray pulsar candidates by a Random Forest classification of unassociated sources from the 4FGL catalogue. Each source was observed for 10 minutes on two separate epochs using MeerKAT's L-band receiver (856--1712\,MHz), with typical pulsed flux density sensitivities of $\sim$100$\,\upmu$Jy. Nine new MSPs were discovered, eight of which are in binary systems, including two eclipsing redbacks and one system, PSR~J1526$-$2744, that appears to have a white dwarf companion in an unusually compact 5\,hr orbit. We obtained phase-connected timing solutions for two of these MSPs, enabling the detection of gamma-ray pulsations in the \textit{Fermi}-LAT data. A follow-up search for continuous gravitational waves from PSR J1526$-$2744 in Advanced LIGO data using the resulting \textit{Fermi}-LAT timing ephemeris yielded no detection, but sets an upper limit on the neutron star ellipticity of $2.45\times10^{-8}$. We
  also detected X-ray emission from the redback PSR~J1803$-$6707 in data from the first eROSITA all-sky survey, likely due to emission from an intra-binary shock.
  \end{abstract}


\clearpage
\section{Introduction}
Observations by the Large Area Telescope \citep[LAT,][]{LAT} onboard the \textit{Fermi Gamma-ray Space Telescope} have led to the detection of gamma-ray pulsations from nearly 300 pulsars\footnote{\url{http://tinyurl.com/fermipulsars}} (The Fermi-LAT Collaboration, 2023, in prep.). These fall into two main classes: canonical pulsars that are still rapidly spinning-down from their initial rotation periods; and millisecond pulsars (MSPs) that have been spun-up (or ``recycled'' \citealt{Bhattacharya+1991}) to rates of hundreds of rotations per second by accretion from an orbiting star \citep{Smarr1976,Alpar+1982}. 

Both pulsar classes have characteristic gamma-ray emission properties (curved spectra and low flux variability over time) that are distinct from those of other gamma-ray emitting objects \citep[e.g.,][]{Ackermann2012+classification}. These properties can be used to identify promising pulsar-like gamma-ray sources, which can then be targeted with radio telescopes to detect pulsations and confirm their nature \citep{Ray2012+PSC}. The few-arcmin localisation regions of unassociated \textit{Fermi}-LAT sources 
enable targeted, long and repeated observations of promising sources, and a higher detection efficiency than can be achieved when surveying a broad region of the sky.

This method has proven highly successful at discovering new MSPs; more than a quarter of the 400 MSPs known in the Galactic field have been discovered in \textit{Fermi}-LAT sources\footnote{\label{GalMSPsfootnote}\url{http://astro.phys.wvu.edu/GalacticMSPs/}} \citep[e.g.,][]{Ransom2011+3MSPs,Cognard2011+Nancay,Keith2011+Parkes,Kerr2012+Parkes,Barr2013+Effelsberg,Camilo2015+Parkes, Cromartie2016+Arecibo} in a global effort coordinated by the \textit{Fermi} Pulsar Search Consortium \citep{Ray2012+PSC}.
These searches have been particularly effective in finding exotic but elusive ``spider'' binary pulsars \citep[``black widows'' and ``redbacks'',][]{Roberts2013+Spiders} whose long radio eclipses due to diffuse intra-binary material make them easily missed in single-pass untargeted surveys. Around two-thirds of the known spider binaries in the Galactic field were found by targeting \textit{Fermi} sources\footnotemark[\value{footnote}].

Many new MSPs remain to be found amongst the LAT sources: pulsars make up around 6\% of the identified or associated\footnote{Gamma-ray sources are only deemed ``identified'' if they have pulsed gamma-ray emission, gamma-ray variability that correlates with that seen in other wavelengths, or are resolved and have an angular extent consistent with that of a known source seen at other wavelengths. If a known source is likely to be the source of the gamma-ray emission, but these conditions are not met, then the gamma-ray source is described as ``associated'' but not identified.} sources
in the recent 12-yr iteration \citep[][hereafter \citetalias{4FGL-DR3}]{4FGL-DR3} of the \textit{Fermi}-LAT Fourth Source Catalog \citep[][hereafter \citetalias{4FGL}]{4FGL}, while over 2000 sources remain unassociated. 
Observed gamma-ray and radio fluxes from pulsars are not strongly correlated with one another \citep[][The Fermi-LAT Collaboration, 2023, in prep.]{2PC}, and so prospects remain high for detecting radio pulsars even in faint new gamma-ray sources that have only recently been detected thanks to the exposure that has accumulated during \textit{Fermi}'s ongoing all-sky survey. This also means that new radio MSPs discovered within faint new \textit{Fermi}-LAT sources can still be bright enough to be valuable astrophysical tools, and indeed several new MSPs found using this method have been added to pulsar timing array (PTA) projects aiming to detect gravitational waves \citep[e.g.][]{Spiewak2022+MeerTIME}. 

The potential importance of detecting new MSPs in \textit{Fermi}-LAT sources is illustrated by the fact that several previous discoveries found in this manner now mark the extreme edges of the MSP population, and are therefore the best current probes for several fundamental astrophysics questions. These include: the fastest known Galactic MSP \citep[PSR~J0952$-$0607,][]{Bassa2017+J0952}; the pulsar binary system with the shortest known orbital period \citep[PSR~J1653$-$0158,][]{Nieder2020+J1653}; and a group of massive black-widow MSPs that probe the maximum neutron star mass \citep{Romani2022+J0952}. 

One crucial benefit of finding a new MSP within a gamma-ray source is that gamma-ray pulsations can often be detected and timed directly in the \textit{Fermi}-LAT data. Initial timing solutions can be refined and extrapolated backwards to the start of the LAT data (which currently spans more than 14 years), providing long and precise rotational ephemerides without the need for lengthy radio timing campaigns. Illustrating the potential scientific benefits of this capability, gamma-ray timing of the recently discovered black-widow PSR~J1555-2908 \citep{Ray2022+J1555} may have revealed a second, planetary mass object in a long period orbit around the inner binary system \citep{Nieder2022+J1555}. The LAT data have also even been recently exploited to build a gamma-ray PTA \citep{Kerr2022+FermiPTA}, whose sensitivity to a stochastic gravitational wave background may reach that of current radio PTAs within a decade.
While searches for pulsations in the \textit{Fermi}-LAT data itself can reveal new MSPs \citep[e.g.,][]{Pletsch2012+J1311,Clark2018+EAHMSPs}, trials factors and computational costs limit these searches to the brighter pulsars. The detection of gamma-ray pulsations from MSPs in binary systems (which most are) is also impossible without prior orbital constraints \citep{Nieder2020+Methods}. Radio surveys and initial timing therefore remain critical for expanding the population of Galactic MSPs. 

The efforts of searching for new MSPs in \textit{Fermi}-LAT sources have recently been bolstered by new radio telescopes. These bring capabilities of observing in new parameter spaces \citep[e.g. at low radio frequencies with LOFAR and GMRT,][]{Bassa2017+J0952,Pleunis2017+LOFAR,2021ApJ...910..160B} or with greater sensitivity \cite[e.g. FAST, ][]{Wang2021+J0318}.

In this paper, we present the first results from one such new survey of unassociated \textit{Fermi}-LAT sources, using the MeerKAT radio telescope \citep{Jonas2009+MeerKAT,Jonas2016+MeerKAT}. The full MeerKAT array is around five times more sensitive than the Murriyang Parkes telescope  \citep{Bailes2020+MeerTIME}, the next most sensitive radio telescope in the Southern Hemisphere. The Transients and Pulsars with MeerKAT (TRAPUM) project is a large survey project using MeerKAT to search for new pulsars \citep{Stappers2016+TRAPUM}. All TRAPUM observations target sky locations in which pulsars are particularly likely to lie: globular clusters \citep{Ridolf2021+TRAPUMGCs}; nearby galaxies \citep{Carli2022+SMC}; supernova remnants, pulsar wind nebulae and other TeV sources; and GeV gamma-ray sources. A separate dedicated L-band survey of the Galactic plane (MMGPS-L) is also ongoing using the same instrumentation and processing infrastructure \citep[][Padmanabh et al., 2023, in prep.]{Kramer2016}. To date, the TRAPUM and MMGPS-L searches have discovered more than 150 new pulsars\footnote{\url{http://trapum.org/discoveries/}}, the majority of which are MSPs. Here we present the first results from TRAPUM's survey of unassociated \textit{Fermi}-LAT sources, which led to 9 of these MSP discoveries.

The paper is organised as follows: Section~\ref{s:survey} describes the survey setup (recording and processing infrastructure, target selection and observation strategy); Section~\ref{s:results} presents the new discoveries and subsequent investigation (localisation, timing and multi-wavelength follow-ups) and an estimate of the survey's sensitivity; and finally a brief discussion and conclusions follow in Sections~\ref{s:discussion} and \ref{s:conclusions}. 

\section{Survey Properties}
  \label{s:survey}
\subsection{MeerKAT and the TRAPUM Processing Infrastructure}
MeerKAT is a radio interferometer located in the Karoo, South Africa, consisting of 64 antennas with 13.5-m effective diameter. Here we give a brief description of MeerKAT and the TRAPUM infrastructure used to perform our pulsar search observations. For a full technical description of the instrument, we refer the reader to \citet{Jonas2016+MeerKAT}. At the time of data taking (between 2020 June and 2021 February), two receivers were available: the L-band receiver operating between 856--1712~MHz and the Ultra High Frequency (UHF) receiver between 544--1088~MHz. This survey was conducted at L-band, but follow-up observations of new pulsars were also made at UHF.

At the L-band centre frequency, a coherent tied-array beam produced using all 64 MeerKAT antennas has a typical full-width at half maximum on the order of a few arcseconds. Furthermore, the data rate of complex voltages from the antennas is too high to record while observing. Efficiently searching the several-arcmin localisation region of an unassociated \textit{Fermi}-LAT source in a single pointing therefore relies on the ability to form and record the Stokes intensities from a large number of coherent beams simultaneously. This capability is provided by the Filterbanking BeamFormer User Supplied Equipment (FBFUSE), a 32-node, GPU-based, software beamformer developed by the Max Planck Institute for Radio Astronomy \citep{Barr2018+FBFUSE, Chen2021+Beamformer}. FBFUSE coherently sums the channelised complex voltages in real-time, using sky position-dependent complex weights computed by the purpose-built \texttt{Mosaic} software\footnote{\url{https://github.com/wchenastro/Mosaic}} from a delay model provided by the MeerKAT Science Data Pipeline. FBFUSE can produce channelised time series for up to 864 coherent beams, as well as one incoherent beam produced by summing the (real-valued) Stokes intensities from each antenna. FBFUSE's beamforming algorithm requires a multiple of four antennas, and not all antennas are available for all observations. As such, we used either 56 or 60 antennas during our observations, depending on availability. 

The channelised data from each coherent beam are then recorded onto a distributed 3.5-PiB file system accessible from the Accelerated Pulsar Search User Supplied Equipment (APSUSE) instrument, a second, 60-node computer cluster, with two NVIDIA GeForce GTX 1080 Ti GPUs per node, on which the pulsar search takes place. The number of beams that can be stored is limited by the data rate at which APSUSE can record, and so down-sampling in time from the native data rate is necessary for a large number of beams to be recorded. All of our observations used the 4096-channel MeerKAT F-engine channeliser mode, but beamformed spectra were down-sampled in time by a factor of 16 from the native time resolution ($4096/856\,{\rm MHz} = 4.785\,\upmu$s) to give 76-$\upmu$s time resolution. With these time and frequency resolutions, up to 288 coherent beams could be formed and recorded. Additionally downsampling in frequency by a factor of two allowed for 480 coherent beams with 2048 frequency channels. Both of these recording modes were used in our survey. 

The full set of filterbank files for 480 coherent beams constitute 46 TiB of data per hour of observation, allowing at most 73 hours of data in this format to be stored for processing on APSUSE. It was therefore necessary to process these data quickly, identify promising candidates for further follow-up, and delete the raw data to ensure there was sufficient storage space for other TRAPUM projects to continue observing. Only the filterbank files for beams in which promising candidates were identified were retained for later use.

\subsection{Observing strategy}
The very high gain and low system temperature of the MeerKAT array ($G = 2.8$~K~Jy$^{-1}$ for the full array and $T_{\rm sys} = 18$~K, \citealt{Bailes2020+MeerTIME}) enable the detection of pulsars with low flux densities, even with short observation lengths. This motivated a strategy involving short pointings towards as many sources as possible. We chose 10 minute observations, for which we still obtain flux density limit estimates of around $95\upmu$Jy that compare favourably against previous surveys of Southern sources with longer observations (see Sections~\ref{s:sensitivity} and \ref{s:discussion}).

Previous surveys for MSPs within unidentified \textit{Fermi}-LAT sources have revealed the importance of observing sources more than once to mitigate non-detections due to scintillation \citep{Camilo2015+Parkes} or unfavourable orbital phases (e.g. due to spider eclipses or ``jerk'' effects \citealt{Andersen2018+Jerk}). In this paper we describe the first two passes of this survey, both performed at L-band. At least two further passes are planned for each source at the UHF band, where the lower frequency will provide additional sensitivity to pulsars with steep spectra, but where propagation effects and dispersive smearing are larger.

During each observation, FBFUSE can be configured to form coherent beams that are distributed within the primary field-of-view either at pre-specified locations, or automatically using an optimal hexagonal tiling pattern. In this latter mode, used for all of our survey observations, the beam spacing is defined by an overlap parameter, which is the fractional sensitivity level of a 2-dimensional Gaussian function fit to the tied-array beam response, simulated by \texttt{Mosaic} \citep[see][]{Chen2021+Beamformer} at the centre of the frequency band, at the point midway between two neighbouring beams. The simple Gaussian model of the earlier versions of \texttt{Mosaic}, used for all our survey observations, tended to overestimate the true overlap when using the full MeerKAT array, and so e.g. points midway between beams in a tiling with an intended $50\%$ overlap actually only achieved $\sim$40$\%$ sensitivity. 

In the first pass of all our targeted sources, we used the 4096-channel mode, with a maximum of 288 coherent beams. The beam tiling patterns for each source were configured with a desired overlap of $50\%$. We prioritised observing sources with larger positional uncertainties when they were at lower elevations, where the coherent tied-array beams covered a larger solid angle, to ensure that the full localisation region could be covered. For the sources with the largest uncertainty regions, observations were scheduled to ensure a sufficiently low elevation that the beam tiling covered a circle with a radius at least as large as the semi-major axis of the elliptical 95\% confidence region (hereafter $r_{\rm 95\%}$) from \citetalias{4FGL}. No significant sensitivity penalty due to ground spillover is incurred for elevations above $30\degr$. For better-localised sources, the outer beams were well outside the \textit{Fermi} source region. 

\begin{figure}
  \centering
  	\includegraphics[width=\columnwidth]{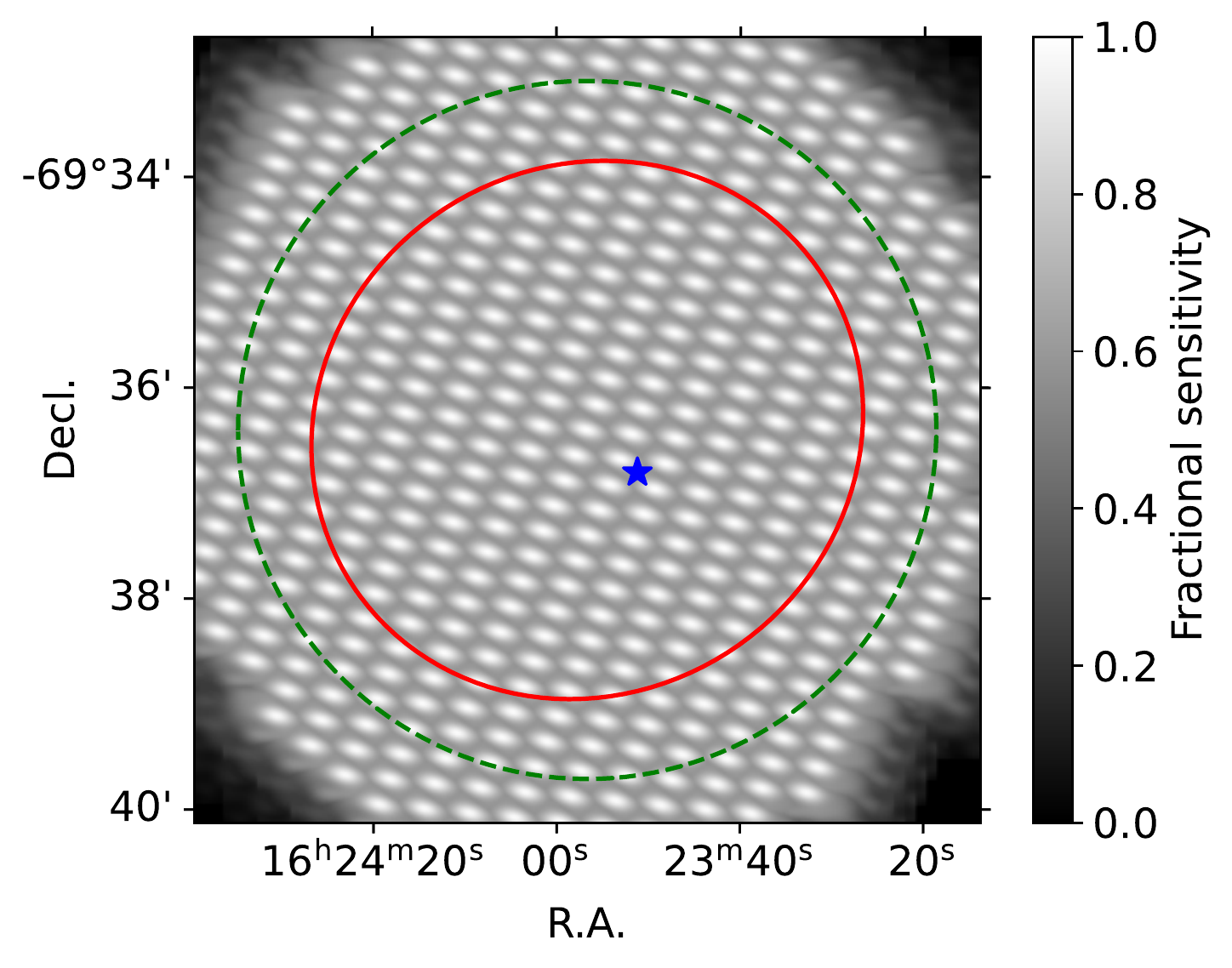}
        \caption{An example of the coherent beam tiling pattern, from the second observation of 4FGL~J1623.9$-$6936. The greyscale image shows the fractional sensitivity according to the \texttt{Mosaic} beam simulation, maximised over neighbouring coherent tied-array beams, reaching 100\% at the centre of each beam. The solid red ellipse is the 95\% confidence region for the \textit{Fermi}-LAT source. The dashed green circle shows the minimum region that we aimed to cover, and has a radius corresponding to the semi-major axis of an approximate 99\% confidence region. The blue star shows the location of PSR~J1623$-$6936 discovered in this observation. The positional uncertainty (obtained in Section~\ref{s:localisation}) is smaller than this marker.}
    \label{f:tiling_example}
\end{figure}

Further development of FBFUSE after our survey began provided the capability to alter the tiling overlap between sources within one observing block. For 75\% of the observations in the second pass, we used the 2048-channel, 480-beam mode and adjusted the tiling overlaps for each source to maximise sensitivity while ensuring that the coherent beams covered a circular region with radius at least $r = \sqrt{\log(0.01)/\log(0.05)} \,r_{\rm 95\%}$, the semi-major axis of an approximate 99\%-confidence ellipse. An example of the resulting tiling pattern is shown in Figure~\ref{f:tiling_example}.
All sources had $r_{\rm 95\%} < 5\arcmin$, such that the tiled region covered a very small patch around the centre of the $\sim $1$\deg$ primary beam, meaning no significant sensitivity loss occurs for coherent beams that are not located at the central pointing position.

\subsection{Search Pipeline}
The filterbank data from all coherent tied-array beams and incoherent beams were searched by a dedicated pipeline built around the  \texttt{peasoup}\footnote{\url{https://github.com/ewanbarr/peasoup}} GPU-accelerated pulsar search code, which performs an FFT-based acceleration search via time-domain resampling with incoherent harmonic summing (\citealt{Barr2020+peasoup}, and described in detail in \citealt{Morello2019+HTRUrepro}). 

The data were de-dispersed at trial dispersion measures (DMs) up to $300$~pc~cm$^{-3}$ with spacing $\Delta\textrm{DM}=0.06$~pc~cm$^{-3}$. This range covers the maximum DMs predicted by the Galactic electron density model of \citet[hereafter \citetalias{YMW16}]{YMW16} in the direction of any of our target sources, and 70 of our 79 target sources have maximum predicted DMs that are less than half of this range (i.e. below 150~pc~cm$^{-3}$).

The acceleration search covered a range of $a \pm 50$~m~s$^{-2}$, which is slightly less than twice as large as the maximum acceleration seen from any known fully-recycled binary MSP ($26.1$~m~s$^{-2}$ from PSR J0024$-$7204V, \citealt{Ridolfi2016+47Tuc}) in the ATNF Pulsar Catalogue \citep[v1.65,][]{psrcat}. Higher accelerations have been observed from relativistic binaries, such as pulsars with massive white dwarf companions or double neutron star systems, but these are all mildly-recycled systems, which typically do not have sufficient spin-down power to emit gamma-ray pulsations, although the Double Pulsar is one notable exception \citep{Guillemot2013+doublePSR}. At each trial DM, the spacing between acceleration trials was chosen according to the scheme described in \citet{Morello2019+HTRUrepro}, which ensures that signals lying between two acceleration trials suffer a total smearing that is not more than 10\% larger than the other (unavoidable) smearing effects due to intra-channel dispersion smearing and finite time resolution. 

After performing the acceleration search, a clustering algorithm was used to search for clusters of candidates from neighbouring coherent beams with similar periods and DMs. Candidates with periods close to known radio-frequency interference (RFI) signals were excluded. This clustering step also aims to distinguish between astrophysical and terrestrial RFI signals, based on how the signal-to-noise ratio (S/N) drops off in neighbouring beams as a function of the angular offset from the beam in which the signal is detected most strongly. With the exception of side-lobe detections of very bright pulsars, signals from astrophysical point sources should only be detected in single beams, or in a small number of beams close to their sky positions. Candidates are therefore rejected if they are detected in many non-neighbouring coherent beams with S/Ns that drop off too slowly with angular offset, as this is indicative of a terrestrial interference signal.

The data from the beam containing the strongest detection of each candidate cluster were then folded using the \texttt{PulsarX} software\footnote{\url{https://github.com/ypmen/PulsarX}}, and folded candidates were scored by the Pulsar Image-based Classification System (PICS) machine-learning classifier \citep{Zhu2014+PICS}. Candidates surpassing a conservatively low PICS-score threshold of $10\%$, typically a few hundred per pointing, were then grouped for visual inspection. As mentioned previously, the full set of raw data could not be stored indefinitely, except for data from beams containing high-confidence pulsar candidates. Reduced data products, i.e. candidate lists, and folded data for every candidate, were retained from all observations.

\subsection{Target Selection}
\label{s:targets}
We built a list of observing targets by identifying pulsar-like unassociated \textit{Fermi}-LAT gamma-ray sources within the \citetalias{4FGL} catalogue.

Several studies have had success in using machine-learning classification techniques to identify pulsar candidates from the population of unassociated sources detected by the \textit{Fermi} LAT \citep{Lee2012+GMM,SazParkinson2016+ML,Luo2020+ML,Finke2021+DNN}. These methods rely on the fact that gamma-ray pulsars have characteristic spectral properties that distinguish them from other gamma-ray emitting source classes: 1) gamma-ray pulsar spectra typically have significant curvature, in that they deviate from a simple power-law spectrum due to a sub-exponential cutoff at photon energies above a few GeV; and 2) gamma-ray pulsars are very stable emitters on long time scales. The Random Forest algorithm \citep{Breiman2001+RF} has been shown to perform well for the purpose of classifying \textit{Fermi}-LAT sources \citep{SazParkinson2016+ML,Luo2020+ML}, and so we employed this method to rank sources from \citetalias{4FGL}.

We used five parameters from \citetalias{4FGL} for the classification: \texttt{PLEC\_SigCurve}, the significance of the log-likelihood improvement when fitting the source spectrum with a (curved) sub-exponentially cutoff power-law spectrum typical of pulsars, compared to a simple power-law; \texttt{Variability2\_Index}, the chi-squared value of the energy flux measured in 2-month time bins; \texttt{Signif\_Avg}, the point-source significance (brighter sources can have higher curvature and variability significance, including this parameter in the ranking accounts for this); \texttt{PLEC\_Index} and \texttt{PLEC\_Expfactor}, the photon index and the pre-factor $a$ in the exponential cutoff term $\exp(-aE^b)$, where $b=2/3$ was used for all unassociated sources and all but six bright pulsars in \citetalias{4FGL}. These final two parameters encode the energy at which the gamma-ray spectrum peaks, which for pulsars typically lies between $0.5-4$~GeV, but which can be at much higher energies for active galactic nuclei (AGN) with curved spectra.

We combined the gamma-ray source classes listed in \citetalias{4FGL} into three broad categories: AGN, pulsars, or ``other'' (which contains e.g. supernova remnants and pulsar wind nebulae). The classifier was trained to identify these classes using sources with highly likely or confirmed associations listed in \citetalias{4FGL}. To evaluate the classifier performance, we removed 33\% of the \citetalias{4FGL} sources, chosen at random, trained the classifier on the remaining population, and compared the classifier results to the known association classes. For identifying pulsars, the classifier had a precision of 82\% (i.e., 82\% of sources predicted to be pulsars were in fact pulsars) and a recall of 71\% (i.e. 71\% of pulsars in the sample were correctly identified as such).

For each unassociated source in \citetalias{4FGL} the Random Forest algorithm estimates the probability, $P({\rm psr})$, of this source being a pulsar. We used this list of unassociated sources, ranked by their predicted pulsar probability, and made further cuts to reduce the number of sources to observe. First, only sources with declinations below $+20^{\circ}$ were included. Next, we removed sources with Galactic latitudes within $\left|b\right| < 10^{\circ}$ of the Galactic plane. This is because gamma-ray pulsars close to the Galactic plane tend to be slowly-spinning young pulsars, which have much narrower radio beams, and are therefore more likely to be radio-quiet gamma-ray pulsars undetectable to our survey. The Galactic plane is also being surveyed with MeerKAT with similar sensitivity as part of the MMGPS-L survey up to $\left|b\right|<5^{\circ}$. Our survey was planned before the MMGPS-L survey region had been finalised, and hence our Galactic latitude cut aimed to avoid redundant observations. We then removed sources whose \citetalias{4FGL} 95\% confidence regions had semi-major axes larger than 5\,arcmin, which is the largest region that could be covered in a single pointing using 288 coherent beams overlapping at 50\% sensitivity.
This semi-major axis cut removes around 40 sources that pass our other cuts, but several of these will be covered in future UHF observations, where the coherent beams are wider. 

Finally, we removed any sources with $P(\rm psr) < 12\%$, which leaves 79 sources to search, while retaining 95\% of the total sum of all $P(\rm psr)$ values. Summing the classifier probabilities gives a (crude) prediction that 38 pulsars (not necessarily all detectable in radio surveys) should exist within these sources, with only 2 expected pulsars in the remaining sources that did not pass the probability threshold. The sources that were searched, along with the pulsar probabilities predicted by the classifier, are listed in Table~\ref{t:targets}. 

\begin{table*}
  \centering
  \footnotesize
  \caption{List of survey observations of \textit{Fermi}-LAT sources. $P({\rm psr})$ is the probability predicted by the Random Forest classifier of each source being a gamma-ray pulsar (see Section~\ref{s:targets}). $S_{\rm opt}$, $S_{\rm 50}$ and $S_{\rm 95}$ are sensitivity estimates (see Section~\ref{s:sensitivity}) from the radiometer equation for a 500\,Hz pulsar, with $\textrm{DM}=100\,\textrm{pc\,cm}^{-3}$ and a 15\% duty cycle, based on optimistic (no losses due to finite frequency, acceleration and DM trial spacings, and for a pulsar lying at the centre of a coherent beam), realistic (median losses) and conservative (95th centile losses) assumptions, respectively.}
  \label{t:targets}
  \begin{tabular}{lccccccccc}
    \hline
    4FGL Source & $P({\rm psr})$ & Epoch 1 & $S_{\rm opt}$ ($\upmu$Jy) & $S_{\rm 50}$ ($\upmu$Jy) & $S_{\rm 95}$ ($\upmu$Jy) & Epoch 2 & $S_{\rm opt}$ ($\upmu$Jy) & $S_{\rm 50}$ ($\upmu$Jy) & $S_{\rm 95}$ ($\upmu$Jy)\\
    \hline

J0048.6$-$6347 & $0.53$ & 59034.2131 & 53 & 114 & 135 & 59196.6117 & 50 & 88 & 97\\
J0139.5$-$2228 & $0.29$ & 59123.8042 & 53 & 110 & 142 & 59196.6191 & 50 & 94 & 112\\
J0251.1$-$1830 & $0.13$ & 59034.206 & 52 & 112 & 136 & 59202.7276 & 49 & 78 & 89\\
J0312.1$-$0921$^{\dagger}$ & $0.94$ & 59034.1976 & 52 & 112 & 132 & 59202.7204 & 49 & 86 & 95\\
J0414.7$-$4300 & $0.45$ & 59034.1435 & 52 & 113 & 136 & 59202.6874 & 49 & 92 & 106\\
J0529.9$-$0224 & $0.3$ & 59065.4275 & 49 & 120 & 144 & 59188.1007 & 51 & 97 & 117\\
J0540.0$-$7552$^{\ddagger}$ & $0.19$ & 59034.221 & 52 & 113 & 138 & 59188.1084 & 50 & 89 & 105\\
J0657.4$-$4658 & $0.26$ & 59065.4594 & 50 & 121 & 142 & 59188.1538 & 50 & 96 & 115\\
J0712.0$-$6431 & $0.17$ & 59065.5499 & 50 & 122 & 189 & 59188.161 & 49 & 99 & 118\\
J0802.1$-$5612$^{\dagger}$ & $0.91$ & 59065.444 & 52 & 128 & 148 & 59182.2759 & 54 & 94 & 108\\
J0940.3$-$7610$^{\ddagger}$ & $0.94$ & 59065.4674 & 51 & 126 & 144 & 59188.1155 & 50 & 81 & 93\\
J0953.6$-$1509 & $0.89$ & 59065.4514 & 49 & 118 & 137 & 59182.2831 & 53 & 94 & 112\\
J1036.6$-$4349$^{\ast}$ & $0.45$ & 59065.4908 & 53 & 130 & 148 & 59182.3887 & 54 & 94 & 112\\
J1106.7$-$1742 & $0.57$ & 59065.498 & 50 & 122 & 140 & 59182.3508 & 53 & 94 & 113\\
J1120.0$-$2204$^{\ddagger}$ & $0.75$ & 59065.5058 & 52 & 126 & 143 & 59188.1761 & 49 & 64 & 74\\
J1126.0$-$5007 & $0.21$ & 59065.4357 & 50 & 119 & 224 & 59182.412 & 54 & 95 & 111\\
J1204.5$-$5032 & $0.5$ & 59020.8389 & 50 & 113 & 182 & 59188.0925 & 51 & 99 & 115\\
J1207.4$-$4536 & $0.7$ & 59065.4835 & 52 & 127 & 152 & 59250.2806 & 54 & 103 & 126\\
J1213.9$-$4416 & $0.23$ & 59065.5129 & 51 & 122 & 140 & 59188.1316 & 50 & 87 & 99\\
J1231.6$-$5116$^{\dagger}$ & $0.96$ & 59065.528 & 53 & 130 & 152 & --- & --- & --- & ---\\
J1303.1$-$4714 & $0.15$ & 59065.4207 & 67 & 159 & 248 & 59188.1244 & 51 & 103 & 121\\
J1345.9$-$2612 & $0.34$ & 59065.4747 & 49 & 116 & 133 & 59188.1389 & 51 & 94 & 113\\
J1400.0$-$2415 & $0.91$ & 59020.7561 & 50 & 117 & 134 & 59188.1461 & 51 & 87 & 100\\
J1416.7$-$5023 & $0.55$ & 59020.7634 & 53 & 125 & 147 & 59188.1686 & 54 & 95 & 108\\
J1450.8$-$1424 & $0.14$ & 59065.5427 & 52 & 130 & 158 & 59188.1981 & 51 & 96 & 111\\
J1458.8$-$2120 & $0.46$ & 59065.5578 & 51 & 122 & 139 & 59188.1909 & 51 & 83 & 95\\
J1513.7$-$1519 & $0.19$ & 59020.7276 & 51 & 120 & 143 & 59188.2053 & 52 & 91 & 103\\
J1517.7$-$4446 & $0.24$ & 59065.5205 & 57 & 143 & 168 & 59188.1837 & 55 & 109 & 130\\
J1526.6$-$2743$^{\ast}$ & $0.68$ & 59020.7346 & 51 & 120 & 140 & --- & --- & --- & ---\\
J1526.6$-$3810 & $0.23$ & 59065.5353 & 57 & 140 & 168 & 59188.2124 & 55 & 110 & 132\\
J1539.4$-$3323 & $0.9$ & 59020.7418 & 52 & 116 & 140 & 59182.3427 & 57 & 100 & 116\\
J1543.6$-$0244 & $0.83$ & 59182.3063 & 54 & 94 & 109 & 59250.3519 & 55 & 104 & 125\\
J1544.2$-$2554 & $0.59$ & 59020.749 & 51 & 120 & 138 & 59196.4976 & 52 & 86 & 97\\
J1612.1$+$1407 & $0.84$ & 59020.7713 & 52 & 123 & 144 & 59196.4842 & 66 & 114 & 129\\
J1622.2$-$7202 & $0.37$ & 59065.5655 & 53 & 130 & 153 & 59182.2916 & 54 & 95 & 112\\
J1623.9$-$6936$^{\ast}$ & $0.17$ & 59020.7123 & 51 & 119 & 137 & 59196.5798 & 51 & 83 & 94\\
J1630.1$-$1049 & $0.57$ & 59020.8087 & 51 & 122 & 166 & 59182.2992 & 56 & 98 & 114\\
J1646.7$-$2154 & $0.17$ & 59020.7198 & 53 & 127 & 175 & 59182.3136 & 58 & 102 & 270\\
J1656.4$-$0410 & $0.18$ & 59123.674 & 134 & 271 & 337 & 59182.3591 & 55 & 97 & 112\\
J1659.0$-$0140 & $0.61$ & 59123.7308 & 57 & 116 & 146 & 59182.3735 & 56 & 99 & 114\\
J1709.4$-$0328$^{\ast}$ & $0.13$ & 59123.7082 & 57 & 117 & 147 & 59196.5426 & 53 & 98 & 112\\
J1711.9$-$1922 & $0.35$ & 59020.8238 & 54 & 127 & 154 & 59196.5498 & 54 & 90 & 101\\
J1717.5$-$5804 & $0.66$ & 59020.7791 & 54 & 126 & 147 & 59196.603 & 55 & 92 & 103\\
J1720.6$+$0708 & $0.6$ & 59020.7944 & 53 & 124 & 144 & 59196.521 & 54 & 97 & 112\\
J1722.8$-$0418 & $0.29$ & 59123.7452 & 58 & 118 & 147 & 59196.557 & 53 & 85 & 96\\
J1727.4$+$0326 & $0.13$ & 59182.3663 & 57 & 99 & 115 & 59250.3446 & 57 & 119 & 138\\
J1730.4$-$0359 & $0.86$ & 59123.738 & 58 & 118 & 150 & 59196.5641 & 54 & 82 & 95\\
J1735.3$-$0717 & $0.32$ & 59123.7816 & 59 & 119 & 147 & 59196.5956 & 55 & 99 & 112\\
J1747.6$+$0324 & $0.46$ & 59123.7673 & 59 & 121 & 152 & 59182.3969 & 59 & 103 & 122\\
J1749.8$-$0303 & $0.22$ & 59020.8015 & 55 & 128 & 146 & 59196.5884 & 56 & 95 & 105\\
J1757.7$-$6032$^{\ast}$ & $0.84$ & 59020.8314 & 52 & 121 & 140 & 59182.321 & 56 & 100 & 120\\
J1803.1$-$6708$^{\ast}$ & $0.23$ & 59020.8161 & 51 & 118 & 138 & --- & --- & --- & ---\\
J1813.7$-$6846 & $0.12$ & 59123.6858 & 56 & 115 & 146 & 59182.3281 & 55 & 98 & 118\\
J1816.4$-$6405 & $0.47$ & 59123.6931 & 57 & 116 & 148 & 59182.3353 & 56 & 99 & 120\\
J1816.7$+$1749 & $0.2$ & 59123.7009 & 58 & 118 & 147 & 59196.5063 & 53 & 94 & 105\\
J1818.6$+$1316 & $0.77$ & 59123.7601 & 57 & 118 & 147 & 59196.5282 & 53 & 92 & 104\\
J1822.9$-$4718 & $0.2$ & 59123.7159 & 56 & 117 & 145 & 59202.6346 & 55 & 94 & 104\\
J1823.8$-$3544$^{\ast}$ & $0.13$ & 59034.0991 & 58 & 125 & 150 & --- & --- & --- & ---\\

    \hline
    \multicolumn{10}{l}{$^{\ast}$ New MSP discovered in this work.}\\
    \multicolumn{10}{l}{$^{\dagger}$ Identified as a gamma-ray pulsar by other groups after this work began.}\\
    \multicolumn{10}{l}{$^{\ddagger}$ Identified as a likely pulsar binary system through optical/X-ray observations, but without pulsation detections.}
  \end{tabular}
  \end{table*}

  \begin{table*}
    \centering
    \footnotesize
    \contcaption{List of survey observations of \textit{Fermi}-LAT sources.}
    \label{t:targets_cont}
    \begin{tabular}{lccccccccc}
      \hline
      4FGL Source & $P({\rm psr})$ & Epoch 1 & $S_{\rm opt}$ ($\mu$Jy) & $S_{\rm 50}$ ($\mu$Jy) & $S_{\rm 95}$ ($\mu$Jy) & Epoch 2 & $S_{\rm opt}$ ($\mu$Jy) & $S_{\rm 50}$ ($\mu$Jy) & $S_{\rm 95}$ ($\mu$Jy)\\
      \hline

J1824.2$-$5427 & $0.61$ & 59123.6786 & 55 & 114 & 143 & 59182.3811 & 55 & 96 & 111\\
J1827.5$+$1141 & $0.96$ & 59123.7744 & 59 & 121 & 151 & 59196.5354 & 55 & 92 & 103\\
J1831.1$-$6503 & $0.67$ & 59123.7231 & 55 & 114 & 143 & 59202.6716 & 52 & 73 & 85\\
J1845.8$-$2521 & $0.85$ & 59123.8123 & 61 & 124 & 155 & 59202.6571 & 56 & 96 & 107\\
J1858.3$-$5424$^{\ast}$ & $0.71$ & 59034.1361 & 54 & 118 & 140 & --- & --- & --- & ---\\
J1906.0$-$1718 & $0.33$ & 59034.1066 & 57 & 123 & 150 & 59182.4041 & 57 & 101 & 123\\
J1906.4$-$1757$^{\ast}$ & $0.14$ & 59123.7962 & 58 & 120 & 151 & --- & --- & --- & ---\\
J1913.4$-$1526 & $0.51$ & 59123.789 & 58 & 120 & 150 & 59202.6428 & 54 & 95 & 105\\
J1916.8$-$3025 & $0.59$ & 59123.7525 & 57 & 117 & 147 & 59202.6643 & 53 & 72 & 84\\
J1924.8$-$1035 & $0.78$ & 59202.65 & 54 & 83 & 95 & 59250.31 & 59 & 102 & 128\\
J1947.6$-$1121 & $0.18$ & 59034.1285 & 55 & 118 & 142 & 59202.6791 & 52 & 89 & 100\\
J2026.3$+$1431 & $0.19$ & 59196.5136 & 51 & 90 & 100 & 59250.3282 & 55 & 110 & 130\\
J2043.9$-$4802 & $0.87$ & 59034.1757 & 53 & 116 & 143 & 59202.7498 & 51 & 77 & 88\\
J2112.5$-$3043 & $0.96$ & 59034.183 & 53 & 115 & 138 & 59202.6952 & 50 & 60 & 67\\
J2121.8$-$3412 & $0.23$ & 59034.2283 & 53 & 114 & 141 & 59202.757 & 50 & 94 & 111\\
J2133.1$-$6432 & $0.77$ & 59034.1903 & 53 & 115 & 137 & 59202.7354 & 50 & 81 & 91\\
J2201.0$-$6928 & $0.41$ & 59034.1214 & 53 & 113 & 133 & 59202.7426 & 50 & 86 & 95\\
J2212.4$+$0708 & $0.92$ & 59034.1595 & 53 & 113 & 133 & 59202.7041 & 50 & 87 & 99\\
J2219.7$-$6837 & $0.52$ & 59034.1138 & 53 & 114 & 135 & 59196.4899 & 50 & 78 & 90\\
J2241.4$-$8327 & $0.33$ & 59034.152 & 53 & 113 & 132 & 59202.7119 & 50 & 87 & 96\\
J2355.5$-$6614 & $0.24$ & 59034.1679 & 53 & 115 & 137 & 59196.5724 & 50 & 91 & 104\\
 \hline
    \multicolumn{10}{l}{$^{\ast}$ New MSP discovered in this work.}\\
\end{tabular}
\end{table*}

\section{Results}
\label{s:results}
\begin{figure}
  \centering
  	\includegraphics[width=\columnwidth]{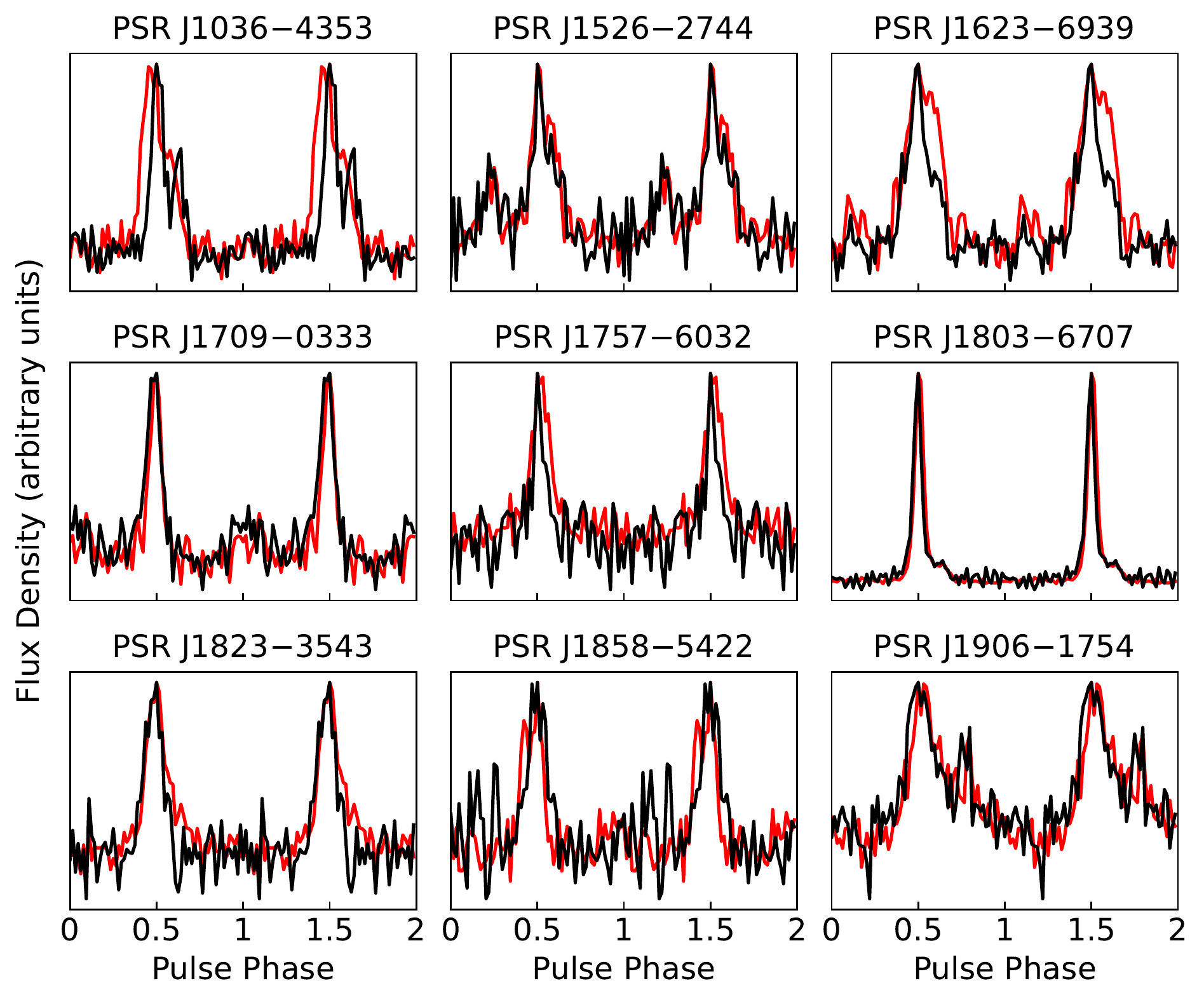}
        \caption{Pulse profiles of the newly-discovered millisecond pulsars. Black and red curves show the profiles from L-band and UHF observations respectively. For clarity, two identical pulses are shown, and pulses have been arbitrarily normalised and phase shifted to peak at phase 0.5. }
    \label{f:pulse_profiles}
\end{figure}

From our two-pass survey, nine candidate signals were identified in the final visual inspection step as likely being new millisecond pulsars, and all of these were later confirmed by additional detections (see Section~\ref{s:followup}). The pulse profiles for the newly-discovered MSPs are shown in Figure~\ref{f:pulse_profiles}. 

Our survey also independently detected PSR~J0312$-$0921, a black-widow MSP discovered recently in a GBT observation of 4FGL~J0312.1$-$0921 (Tabassum et al., 2023, in prep.). Shortly after our first observations, another two target sources, 4FGL~J0802.1$-$5612 and 4FGL~J1231.6$-$5116, were identified as young gamma-ray pulsars by the direct detection of gamma-ray pulsations in the \textit{Fermi}-LAT data by \textit{Einstein@Home}\footnote{\url{https://einsteinathome.org/gammaraypulsar/FGRP1_discoveries.html}}. Neither pulsar was detected in our survey, but this is not surprising, as only a very small fraction of young pulsars discovered in gamma-ray pulsation searches have been detected in radio observations \citep{Wu2018+EatH}.

\begin{table*}
  \centering
  \caption{Properties of newly discovered millisecond pulsars without phase-connected timing solutions. Initial orbital solutions are obtained from fitting sinusoids to observed barycentric spin periods as a function of time. These models are approximate, and so we only provide limited precision on orbital parameters and spin frequency. Positional parameters (R.A. and Decl.) are estimated from the measured signal-to-noise ratios in neighbouring coherent beams using \texttt{SeeKAT} (see Section \ref{s:localisation}), except for J1036$-$4353 whose position is from \textit{Gaia} DR3.}
  \label{t:pulsars}
  \begin{tabular}{lcccc}
    \hline
    Parameter & PSR~J1036$-$4353 & PSR~J1623$-$6939 & PSR~J1709$-$0333 & PSR~J1757$-$6032 \\
    \hline
    R.A., $\alpha$ (J2000) & $10^h36^m30\fs21513(1)$ & $16^h23^m51\fs41^{+0\fs33}_{-0\fs33}$ & $17^h09^m32\fs79^{+0\fs18}_{-0\fs13}$ & $17^h57^m45\fs53^{+0\fs22}_{-0\fs24}$\vspace{1ex}\\
    Decl., $\delta$ (J2000) & $-43\degr53\arcmin08\farcs7252(2)$ & $-69\degr36\arcmin48\farcs3^{+2\farcs2}_{-1\farcs7}$ & $-03\degr33\arcmin17\farcs{7}^{+6\farcs0}_{-2\farcs0}$ & $-60\degr32\arcmin10\farcs7^{+2\farcs6}_{-2\farcs3}$\vspace{1ex}\\
    Dispersion measure, DM (pc cm$^{-3}$) & 61.1 & 46.4 & 35.7 & 62.9\\
    Spin frequency, $\nu$ (Hz) & 595.2 & 415.0 & 283.8 & 343.3\\
    Orbital Period, $P_{\rm orb}$ (d) & 0.259 & 11.01 & --- & 6.28\\
    Projected semi-major axis, $x$ (lt-s) & 0.665 & 6.73 & --- & 9.62\\
    Epoch of ascending node, $T_{\rm asc}$ (MJD) & 59536.31 & 59192.91 & ---& 59183.40\\
    \hline
    Minimum companion mass, $M_{\rm c,min}$ ($M_{\odot}$) & 0.23 & 0.19 & ---& 0.43\\
    Distance (from \citetalias{YMW16}), $d$ (kpc) & 0.40 & 1.3 & 0.21 & 3.5\\
    \hline
    \hline
    Parameter & PSR~J1823$-$3543 & PSR~J1858$-$5422 & PSR~J1906$-$1754 & ---\\
    \hline
    R.A., $\alpha$ (J2000) & $18^h23^m43\fs06^{+0\fs14}_{-0\fs13}$ & $18^h58^m07\fs92^{+0\fs26}_{-0\fs23}$ & $19^h06^m14\fs94^{+0\fs11}_{-0\fs15}$ & ---\vspace{1ex}\\
    Decl., $\delta$ (J2000) & $-35\degr43\arcmin40\farcs8^{+1\farcs4}_{-2\farcs2}$ & $-54\degr22\arcmin14\farcs6^{+3\farcs8}_{-2\farcs7}$ & $-17\degr54\arcmin33\farcs{7}^{+1\farcs8}_{-3\farcs2}$ & ---\vspace{1ex}\\
    Dispersion measure, DM (pc cm$^{-3}$) & 81.7 & 30.8 & 98.1 & ---\\
    Spin frequency, $\nu$ (Hz) & 421.4 & 424.5 & 347.7 & ---\\
    Orbital Period, $P_{\rm orb}$ (d) & 144.5 & 2.58 & 6.49  & ---\\
    Projected semi-major axis, $x$ (lt-s) & 51.8 & 1.68 & 1.35 & ---\\
    Epoch of ascending node, $T_{\rm asc}$ (MJD) & 59091.36 & 59564.98 & 59304.97 & ---\\
    \hline
    Minimum companion mass, $M_{\rm c,min}$ ($M_{\odot}$) & 0.27 & 0.12 & 0.05 & ---\\
    Distance (from \citetalias{YMW16}), $d$ (kpc) & 3.7 & 1.2 & 6.8 &\\
    \hline
  \end{tabular}
  \vspace{5ex}
\end{table*}

\subsection{Follow-up Observations}
\label{s:followup}
To confirm the pulsar nature of the detected candidates, we performed dedicated follow-up observations with MeerKAT at both L-band and UHF, and checked for archival data from previous search observations of the \textit{Fermi}-LAT sources in which they were found. Three pulsars (J1757$-$6032, J1803$-$6707 and J1823$-$3543) were re-detected in archival data from the Parkes radio telescope.

The five high-confidence pulsar candidates detected in the first pass were removed from the scheduled second survey pass, so that they could be observed in a dedicated set of confirmation observations along with candidates from the second pass. In these confirmation observations we employed a very dense tiling (with $90\%$ overlap) with a smaller number of beams around the location of the coherent beam in which the candidate was detected most strongly in the initial survey observations. This dense tiling ensured high sensitivity for re-detecting these candidates, while also enabling us to more precisely localise each pulsar using the method described below (see~Section~\ref{s:localisation}). 

Eight of the pulsar candidates (excluding PSR~J1036$-$4353, discussed below) were re-detected in these confirmation observations. One pulsar, PSR~J1709$-$033 was only seen in the UHF observation, most likely due to unfavourable scintillation that was seen during observations in which this pulsar was detected.

One pulsar, PSR~J1036$-$4353 was not included in the confirmation observations, as it had not been immediately identified as a candidate due to a bug in the folding pipeline that caused it to be folded with the wrong acceleration sign. Instead, it was identified at a later date when the folded archives were corrected. It was confirmed in a later UHF observation as part of the next stage of this survey, which will be presented in a future paper.

\subsection{Localisation}
\label{s:localisation}
Following the detection of a pulsar in our survey, the sky position could be estimated to a much higher precision than the size of a coherent beam by triangulation using the measured S/Ns in neighbouring beams. The method with which we localised candidates from our survey is described in Bezuidenhout et al. (2022, RASTI, submitted), based on the concept outlined in \citet{Obrocka2015+reloc}, which works as follows. Given the model of the coherent beam point spread function (PSF) provided by \texttt{Mosaic}, one can compute the expected ratio of the S/Ns that would be recovered in two neighbouring beams for a pulsar at any given point nearby. This expected ratio will match the observed ratio, within uncertainties, for a strip of positions between the two beams. These strips can then be computed for each pair of beams, and the location of the candidate can be inferred from where these strips all cross one another.

This procedure is implemented by the \texttt{SeeKAT} package\footnote{\url{https://github.com/BezuidenhoutMC/SeeKAT}}. We used \texttt{SeeKAT} on each of the confirmation observations, which had the most dense beam tilings. The best-fitting positions and uncertainties were in good agreement across the three observations (one at L-band, two at UHF), and so we combined these results by summing together the log-likelihood surfaces from each observation.

\begin{figure}
  \centering
  \includegraphics[width=0.94\columnwidth]{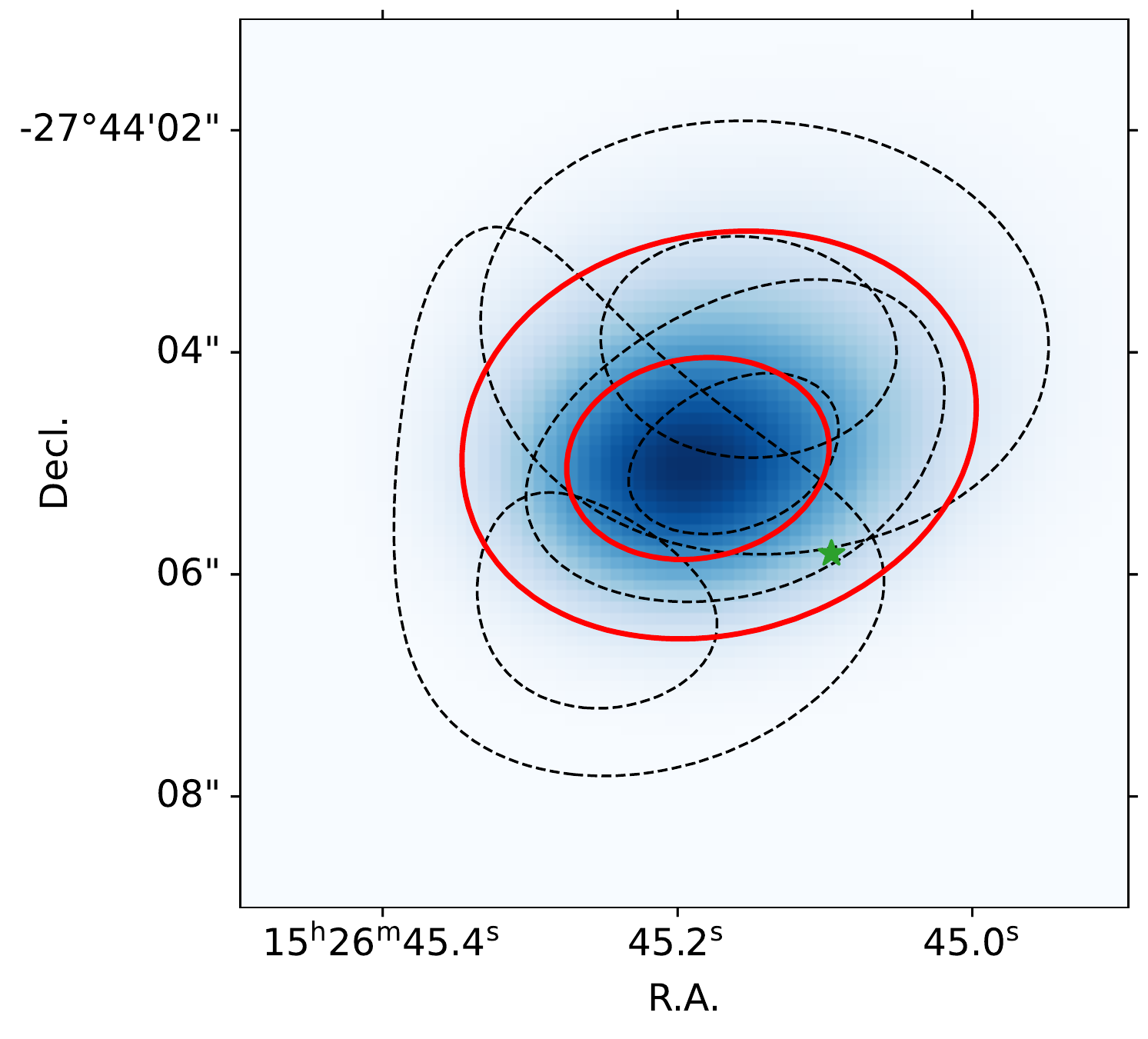}
        \caption{An example of pulsar localisation using \texttt{SeeKAT}, from the three confirmation observations of PSR~J1526$-$2744. The dashed black contours show the localisation probabilities in three separate observations. The colour scale and the solid red contour lines show the joint likelihood (i.e. the product of the likelihoods from all three observations). Contour lines are at approximate 1$\sigma$ and 2$\sigma$ levels. The green star is at the \textit{Fermi} timing position for this pulsar.}
    \label{f:seekat}
\end{figure}

The best-fitting positions and uncertainties for each pulsar are given in Table~\ref{t:pulsars}, and an example of the localisation log-likelihood surface is illustrated in Figure~\ref{f:seekat}. The \texttt{SeeKAT} positional uncertainties for new pulsar candidates are typically on the order of a few arcseconds, sufficiently precise to enable the identification of potential counterparts in multi-wavelength catalogues. 

\subsection{Timing}
Follow-up timing campaigns have begun for all newly detected pulsars. Initially, all timing observations were performed at other telescopes.
The pulsars visible from the Northern hemisphere are followed up mainly using the Nan\c{c}ay (for PSRs~J1526$-$2744 and J1823$-$3543) and Effelsberg (for PSRs~J1709$-$0333 and J1906$-$1754) telescopes at L-band, while all other pulsars (and, initially, also those observed at other telescopes) are followed up at the Murriyang Parkes telescope, using the Ultra-wide-band Low (UWL) receiver \citep{Hobbs2020+UWL}, covering a frequency range from 0.7 to 4~GHz. Depending on the discovery S/N at MeerKAT or, when available, the S/N at the improved position, we follow-up the TRAPUM discoveries for 1 to 2 hrs. Observations are carried out in search mode. Whenever possible the pulsars have been observed with a pseudo-logarithmic cadence to help achieve phase coherence in our timing analysis. Phase-connected timing solutions for two MSPs, described in detail below, were obtained
using the \texttt{Dracula} algorithm\footnote{\url{https://github.com/pfreire163/Dracula}} \citep{Dracula}; this was necessary given the sparsity of the timing data in both cases.

\begin{figure}
  \centering
  \includegraphics[width=\columnwidth]{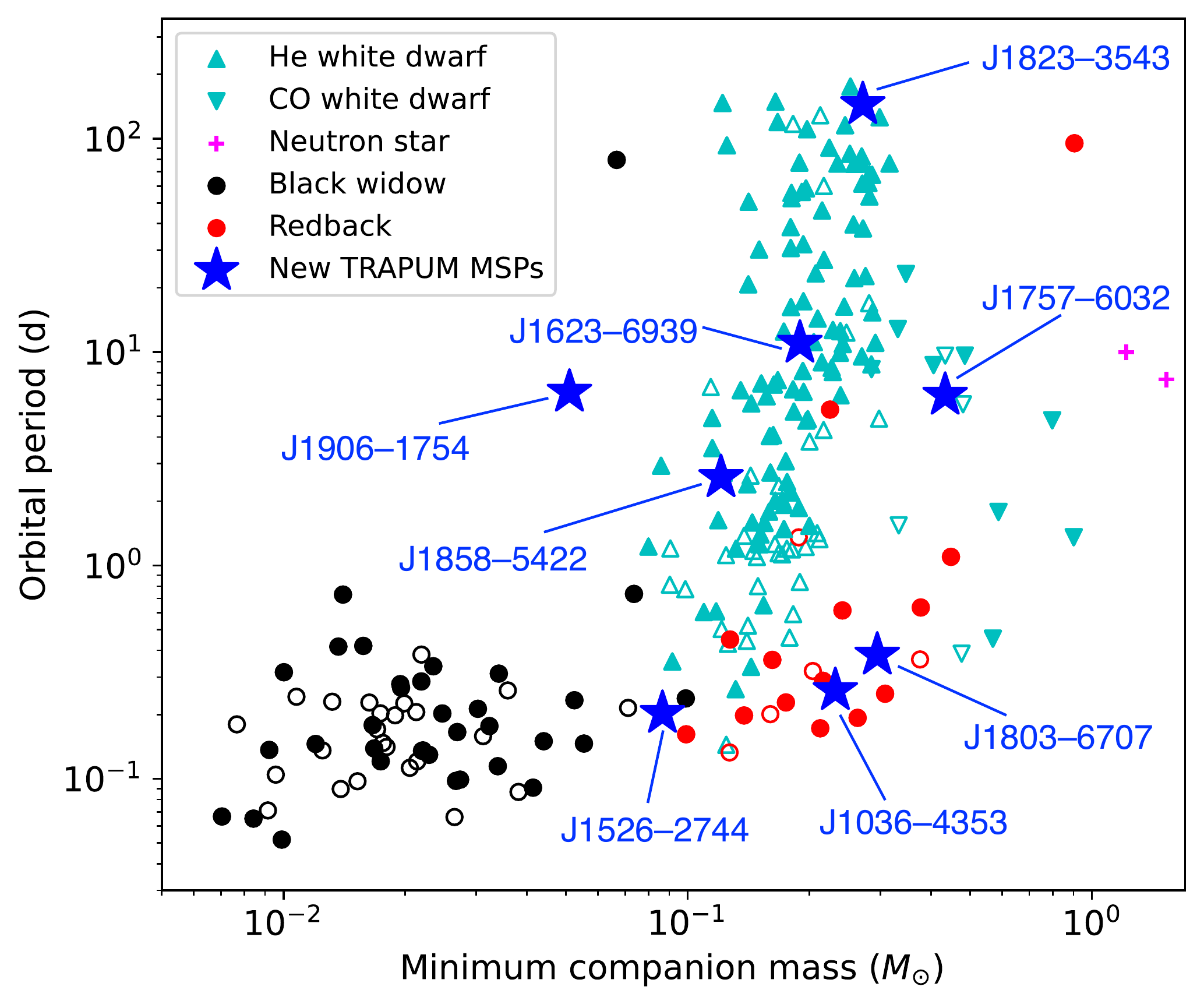}
        \caption{Orbital properties of known binary MSPs. Each companion type, according to the ATNF Pulsar Catalogue \citep{psrcat}, is denoted by a different marker as noted in the legend. We mark two pulsars categorised as having He white dwarf companions in the ATNF Pulsar Catalogue, PSRs~J1518$+$0204C and J1653$-$0158, as black widows following \citet{Pallanca2014+J1518} and \citet{Nieder2020+J1653}, respectively. Filled markers show MSPs that were found by targeting an unidentified \textit{Fermi}-LAT gamma-ray source. The new MSPs from this work are highlighted with blue stars and labelled. }
    \label{f:mass_vs_PB}
\end{figure}

Obtaining a phase-connected timing solution for the remaining seven pulsars has required a dedicated timing campaign using the Pulsar Timing User Supplied Equipment (PTUSE) system developed for the MeerTime project \citep{Bailes2020+MeerTIME}, which additionally provides coherent de-dispersion and full Stokes polarisation information. This MeerKAT timing campaign, along with follow-up searches for gamma-ray pulsations from these seven pulsars, will be presented in a dedicated paper (Burgay et al., 2023, in prep.). For these seven pulsars, we give a preliminary timing solution in Table \ref{t:pulsars}, obtained using \texttt{PRESTO}'s \texttt{fit\_circular\_orbit.py} routine to fit a sinusoidal modulation to the observed barycentric spin periods from multiple observations without requiring phase-alignment across observations. We also plot the orbital properties of the 8 newly-detected binary MSPs (excluding PSR J1709$-$0333 which appears to be an isolated pulsar) according to these ephemerides, in comparison to the overall MSP population, in Figure~\ref{f:mass_vs_PB}.

\subsubsection{PSR J1526$-$2744}
\label{s:J1526}
Using the \texttt{SeeKAT} position as a starting point, we obtained a phase-connected timing solution using TOAs produced from the original L-band observation, the two UHF observations, as well as several Nan\c{c}ay and Parkes observations taken in 2021 as part of our dedicated timing campaign. This timing solution 
provided precise constraints on the pulsar's orbital semi-major axis and time of ascending node, as well as a refined position, but the spin-down rate was not significantly measurable and was highly correlated with the declination.

We used the radio timing solution to search for gamma-ray pulsations in the photon arrival times measured by the \textit{Fermi} LAT. For this we used \texttt{SOURCE}-class gamma-ray photons from within a $3\degr$ region-of-interest around the \texttt{SeeKAT} position, with energies greater than 100\,MeV and with a zenith angle below $105\degr$ according to the ``Pass 8'' \texttt{P8R3\_SOURCE\_V2} \citep{Pass8,Bruel2018+P305} instrument response
functions \footnote{See \url{https://fermi.gsfc.nasa.gov/ssc/data/analysis/LAT_essentials.html}}. To increase sensitivity to faint pulsations, we used \texttt{gtsrcprob} to compute photon probability weights \citep{Kerr2011}, using the spectral and spatial parameters for nearby sources from the \citetalias{4FGL} DR2 catalogue, as well as the \texttt{gll\_iem\_v07.fits} Galactic and \texttt{iso\_P8R3\_SOURCE\_V2\_v1.txt} isotropic diffusion emission models.

Folding the gamma-ray data over the validity interval of the radio timing solution did not yield a significant detection. To search for gamma-ray pulsations in earlier data it was necessary to search over a 5-dimensional parameter space ($\alpha$, $\delta$, $\nu$, $\dot\nu$ and $P_{\rm orb}$). The pulsar's orbital semi-major axis and time of ascending node were already constrained precisely enough by the initial radio ephemeris that only one trial was required in these dimensions. The search was performed using the weighted $H$-test \citep{deJaeger1989+Htest,Kerr2011}, a statistic that normally performs an optimal incoherent sum of the Fourier power in the first 20 harmonics. In this case, we only summed over 3 harmonics, as detecting power in higher harmonics requires increased search grid density in each dimension, but gamma-ray pulsars have most power in lower harmonics. This search detected a significant pulsed signal with $H = 108.9$, and with phase-connected pulsations visible from the start of the LAT data. 

\begin{figure}
  \centering	\includegraphics[width=\columnwidth]{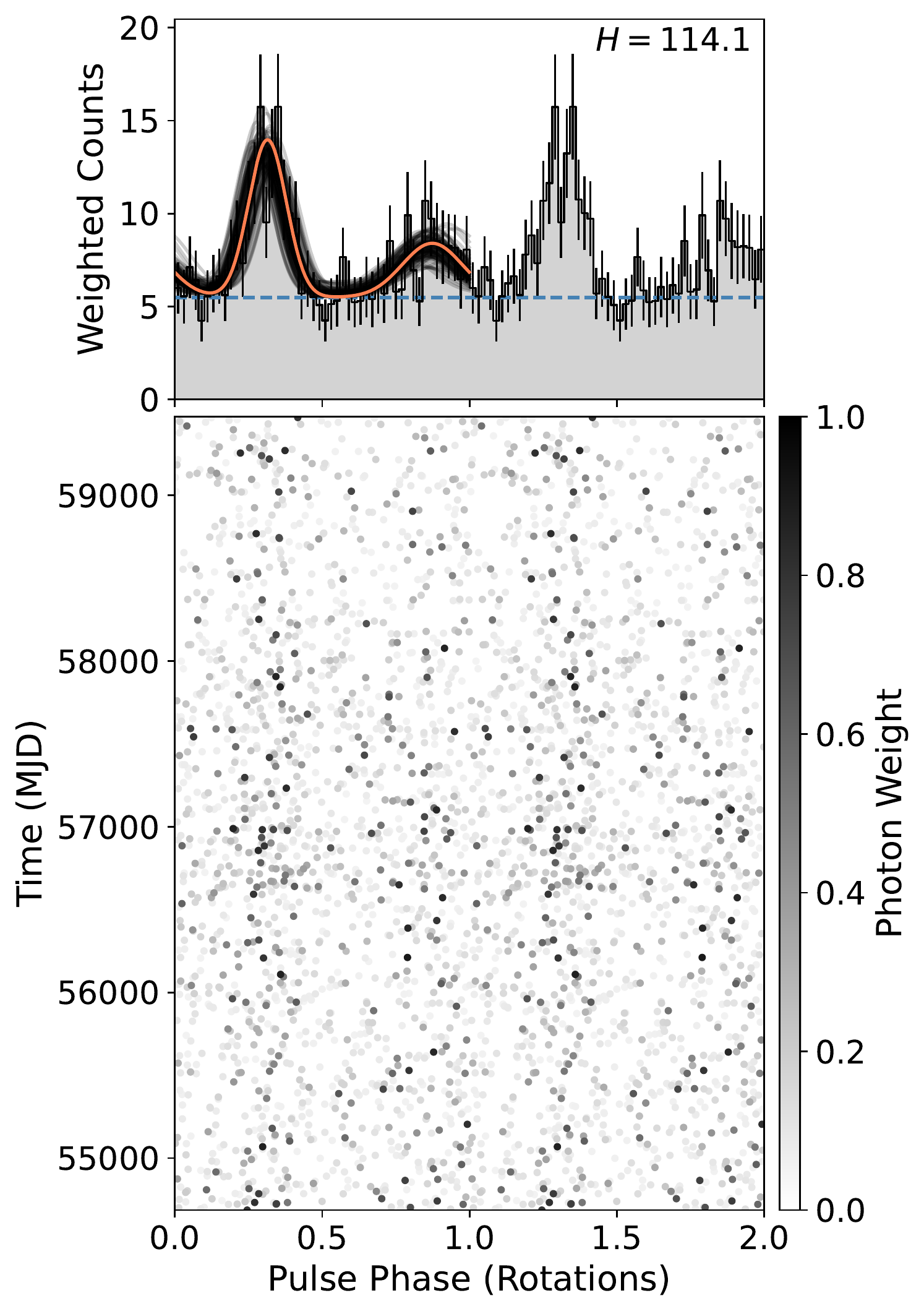}
        \caption{Gamma-ray pulsations from PSR~J1526$-$2744 in \textit{Fermi}-LAT data. The lower panel shows each photon's phase according to the best-fitting timing solution. Each photon is assigned a weight, computed from its incidence angle and energy, and illustrated by greyscale representing the probability that it was emitted by the targeted source as opposed to a fore- or background source. The upper panel shows the integrated pulse profile, computed by summing the photon probability weights in each rotational phase bin. The dashed blue horizontal line shows the estimated background rate.
        In the upper panel, the opaque orange curve shows the best-fitting template pulse profile used to estimate the log-likelihood of the gamma-ray pulsations. The transparent black curves show templates with parameter values randomly drawn from our MCMC samples to illustrate the uncertainty on the pulse profile.
}
    \label{f:J1526_fermi}
\end{figure}

Following this detection, we derived a precise 13-yr gamma-ray timing solution by varying the timing parameters to maximise the unbinned Poisson log-likelihood of the weighted photon phases \citep{2PC} using a template pulse profile consisting of two wrapped Gaussian peaks whose parameters were also free to vary in the fit. The best-fitting parameter values and uncertainties are given in Table~\ref{t:J1526}. The gamma-ray photon phases according to the best-fitting timing model and template pulse are shown in Figure~\ref{f:J1526_fermi}. We also tested for but did not significantly detect proper motion ($\left|\mu\right| < 49\,\textrm{mas\,yr}^{-1}$) or eccentricity ($e < 8\times10^{-4}$). 

\begin{table}
  \centering
  \caption{Timing solution for PSR~J1526$-$2744. Timing parameters are obtained from the gamma-ray timing analysis, with the exception of the DM measurement which is from the original MeerKAT discovery observation. Parameter values are in the Barycentric Dynamical Time (TDB) scale.}
  \label{t:J1526}
  \begin{tabular}{lc}
    \hline
    Parameter & Value \\
    \hline
    \multicolumn{2}{c}{Timing parameters}\\
    \hline
    Data span (MJD) & $54681$--$59476$\\
    Reference epoch (MJD) & 59355.468037\\
    R.A., $\alpha$ (J2000) & $15^h26^m45\fs103(2)$\\
    Decl., $\delta$ (J2000) & $-27\degr44\arcmin05\farcs91(8)$\\
    Dispersion measure, DM (pc cm$^{-3}$) & $30.95(3)$\\
    Spin frequency, $\nu$ (Hz) & $401.7446020975(3)$\\
    Spin-down rate, $\dot{\nu}$ (Hz s$^{-1}$) & $-5.71(1)\times10^{-16}$\\
    Orbital period, $P_{\rm orb}$ (d) & $0.2028108285(7)$\\
    Projected semi-major axis, $x$ (lt s) & $0.22410(3)$\\
    Epoch of ascending node, $T_{\rm asc}$ (MJD) & $59303.20598(1)$\\
    \hline
    \multicolumn{2}{c}{Derived parameters}\\
    \hline
    Spin period, $p$ (ms) & $2.489143587192(2)$\\
    Spin period derivative, $\dot{p}$ & $3.537(6)\times10^{-21}$\\
    Spin-down power, $\dot{E}$ (erg/s) & $9.1\times10^{33}$\\
    Surface magnetic field strength, $B_{\rm S}$ (G) & $9.5\times10^7$\\
    Light-cylinder magnetic field strength, $B_{\rm LC}$ (G) & $5.7\times10^4$\\
    Minimum companion mass, $M_{\rm min}$ ($M_{\odot}$) & 0.083\\
    Distance (from \citetalias{YMW16}), $d$ (kpc) & 1.3\\
    \hline
  \end{tabular}
\end{table}

The nature of the companion star in this system is currently unclear. The short orbital period ($4.9$\,hr) and low minimum companion mass ($0.08\,M_{\odot}$) suggest a heavy black-widow or light redback companion, but no radio eclipses typical of these systems have been seen. Many black-widow binaries have detectable optical counterparts, and so we searched for an optical counterpart to this system using the ULTRACAM
\citep{Dhillon2007+ULTRACAM} high-speed multi-band imager on the 3.5-m New Technology Telescope (NTT) at ESO La Silla. The longest observation lasted 3.5\,hr, covering orbital phases between 0.1--0.8. We did not find any counterpart at the pulsar's timing position, with $3\sigma$ magnitude upper limits of $i_s = 23.2$,$g_s = 24.0$, and $u_s=23.0$ in the deepest 5-minute stacked image, which was obtained close to the companion's superior conjunction, where a heated black-widow companion would appear at its brightest. It therefore seems likely that the companion is a light-weight ($M_{\rm min} = 0.083\,M_{\odot}$) white dwarf. 
If confirmed, e.g. through the detection of a non-variable optical counterpart below our ULTRACAM detection threshold, this would be the shortest orbital period of any known fully-recycled Galactic MSP--WD system, and the second shortest orbit of any PSR--WD binary system, after the relativistic binary PSR~J0348+0432 \citep{Antoniadis2013+J0348}.

For the observed $\textrm{DM}=30.95(3)$\,pc\,cm$^{-3}$, \citetalias{YMW16} predicts a distance of 1.3\,kpc.
The gamma-ray energy flux from this source above 100\,MeV is $G_{100} = 2.5\pm 0.4\times10^{-12}$\,erg\,cm$^{-2}$\,s$^{-1}$, corresponding to a gamma-ray luminosity of $L_{\gamma} = 5.3\times10^{32}\,{\rm erg\,s}^{-1}$. This can be compared to the pulsar's spin-down power $\dot{E} = 4 \pi^2 I \nu \dot{\nu} = 9.1\times10^{33}\,{\rm erg\,s}^{-1}$, for an assumed pulsar moment-of-inertia $I=10^{45}$\,g\,cm$^{2}$,  corresponding to an efficiency $\eta = L_{\gamma} / \dot{E} \approx 6\%$, which is typical of gamma-ray MSPs \citep{2PC}.

Using the timing solution obtained above, we carried out a search for continuous gravitational waves from this pulsar. Continuous gravitational wave emission is expected due to a non-axisymmetric deformation ($\epsilon$) of the neutron star and the dominant signal frequency is twice its rotational frequency. Our search was therefore targeted at a frequency $f$ $\approx$ $\fgw$ Hz and spin down $\dot{f}$ $\approx$ $\fdotgw$ Hz/s and directed at the timing position.

An \textit{a priori} estimate of the detectable strain amplitude using Advanced LIGO data yields a value about an order of magnitude larger than the spin-down upper limit amplitude \citep{2014ApJ...785..119A}, 
\begin{equation}
  h_0^{sd} = \left ( \frac{5}{2}\frac{GI |\dot{\nu}|}{c^{3}d^{2}\nu} \right) ^ {1/2}
    %|\dot{f}_{rot}|}{c^{3}d^{2}f_{rot}} \right) ^ {1/2}
\end{equation}
where we again assume $I=10^{45}$~g~cm$^2$.
This is the strain  amplitude of the signal assuming that all the rotational kinetic energy lost by the pulsar ($\dot{E}$) was converted into gravitational waves. We therefore do not expect a detection, but, in the spirit of leaving no stones unturned {\citep[e.g.,][]{LIGOScientific:2017ous,LIGOScientific:2019mhs,2019ApJ...883...42N,Nieder2020+J1653,2021arXiv211210990T,2022ApJ...935....1A}}, we carried out the search.

We used all of the Advanced LIGO data from the Hanford and Livingston detectors collected during the O1, O2 and O3 runs \citep{LIGOScientific:2019lzm}. The gamma-ray timing solution allows us to coherently combine these datasets using a single template, thus achieving the maximum possible sensitivity this search could have to date. Loud detector glitches in the data were removed through gating \citep{2022PhRvD.105b2005S}. The frequency range relevant to this search is free of known lines in the detectors. The multi-detector, matched-filtering $\oneF$-statistic \citep{2005PhRvD..72f3006C}  was used for the analysis. Our search result has a p-value of \pvalue\%, estimated using off-source data. This result implies a non-detection and based on it we set a 95\% confidence upper limit of $\hULval$ on the intrinsic gravitational wave amplitude. Our upper limit is a factor of $\approx$ $\ulratio$ larger than the spin down upper limit, and the uncertainty on this upper limit is not more than $\hULpercentUncertainty$\% including calibration uncertainties. Translating the amplitude upper limit into an upper limit on the ellipticity of the source, we constrain the ellipticity of J1526$-$2744 to be $<\epUL$, which is close to the minimum ellipticity proposed for millisecond pulsars by \cite{2018ApJ...863L..40W}. 

We also searched in a band of $f$ and $\dot{f}$ accounting for mismatches between the phase of the  gravitational wave signal and the phase locked to the electromagnetic observations. Such mismatches could result from a differential rotation between the parts of the star emitting the gravitational wave and the electromagnetic pulsations or if the star was biaxial and consequently freely-precessing. We searched in a bandwidth of 0.4\% of the spin parameter values following \cite{LIGOScientific:2020gml}, and estimated the p-values of the results as done in \cite{2021ApJ...923...85A}. The results from this band search were also consistent with expectations from Gaussian noise.

\subsubsection{PSR J1803$-$6707}
\label{s:J1803}
PSR~J1803$-$6707 was the first pulsar to be found in our survey, and was quickly confirmed by a detection in 1\,hr of archival Parkes data from 2015. Detection of the pulsar in this observation required a significant jerk term, indicating that the pulsar was likely to be in a short-period binary system.  Pulsations were only detected for $\sim$30~mins in one of two 1\,hr dedicated follow-up observations with Parkes in December 2020, indicative of a wide eclipse typical of black-widow or redback systems.

Establishing an orbital timing solution (given in Table~\ref{t:J1803}) revealed a $\sim$9\,hr orbit and a minimum companion mass of 0.26\,$M_{\odot}$ (assuming a pulsar mass of 1.4\,$M_{\odot}$), as well as significant variations in the orbital period requiring several orbital frequency derivative terms to describe the orbital phase throughout the 1\,yr of phase-connected timing data. These features are characteristic of redback binary systems \citep[e.g.,][]{Deneva2016+J1048}.

\begin{figure}
  \centering
  	\includegraphics[width=\columnwidth]{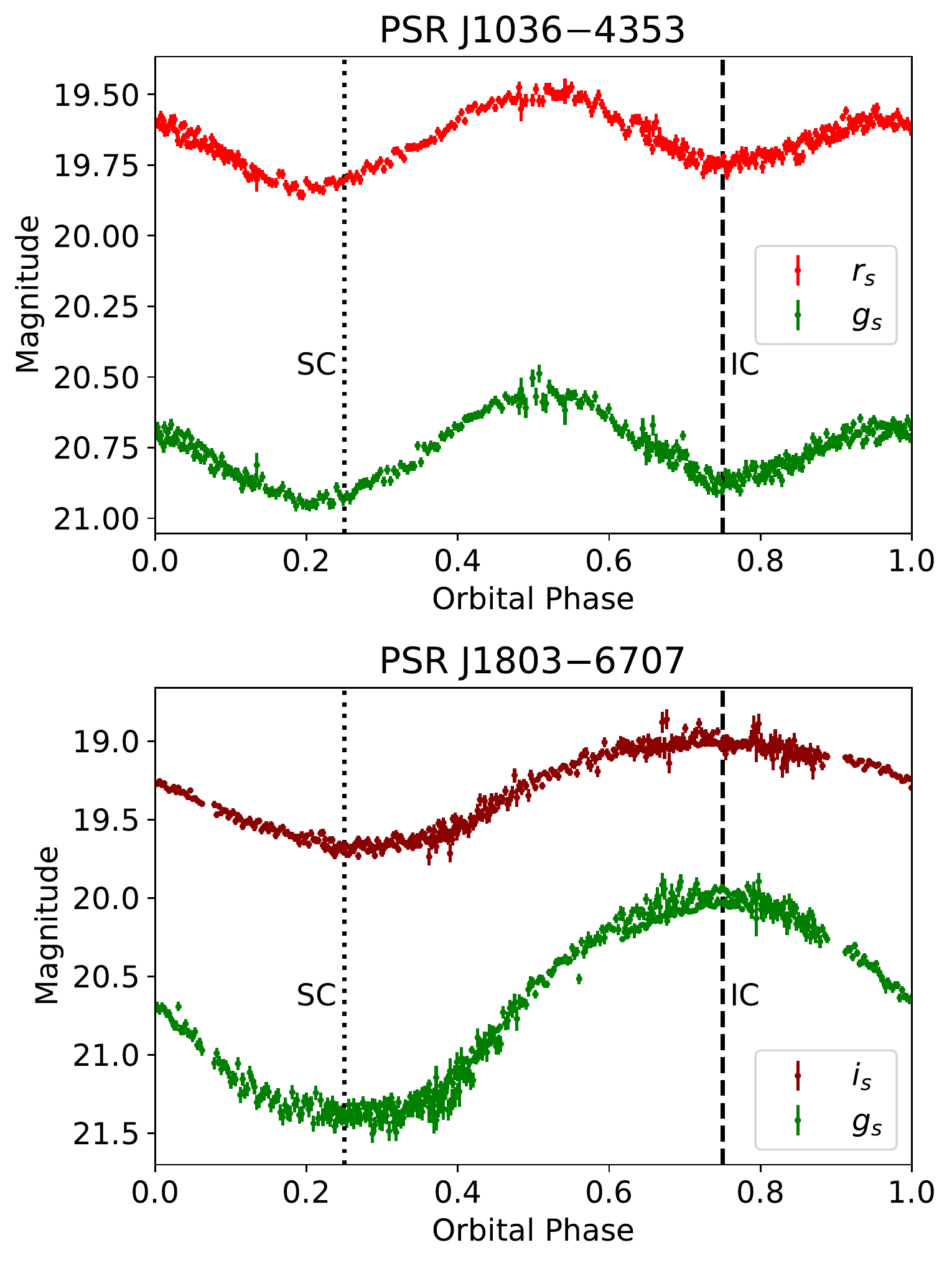}
        \caption{Optical light curves for PSRs~J1036$-$4353 and J1803$-$6707 obtained using ULTRACAM+NTT. Observations are folded on the orbital period using the radio timing ephemerides.  The pulsar's superior and inferior conjunctions are marked with dotted and dashed vertical lines, and labelled SC and IC, respectively. The two peaks in the light curve for PSR~J1036$-$4353 are due to ellipsoidal deformation of the companion star in the pulsar's gravitational field, while the single peak in the light curve for PSR~J1803$-$6707 indicates strong heating from the pulsar. }
    \label{f:ultracam}
\end{figure}

We used the UHF MeerKAT observations to estimate the sky position of this pulsar using \texttt{SeeKAT}. The refined position is coincident with the location of a star in the \textit{Gaia} DR3 catalogue \citep{Gaia,Gaia+DR3}. We also observed this star with ULTRACAM, which revealed optical variability with the same 9\,hr periodicity, confirming that this is indeed the optical counterpart. The optical light curve, shown in Figure~\ref{f:ultracam}, varies by around 1.5\,mag, with a single peak indicative of significant heating via irradiation from the pulsar \citep[similar to e.g., PSR J2215$+$5135,][]{SchroederHalpern2014} but the counterpart is detectable at all orbital phases, as is typical for nearby redback companions which tend to have hot surfaces ($T > 4000$\,K) even on the non-irradiated side. Light curve modelling to estimate properties of the companion such as its temperature, irradiation, radius and the binary inclination angle will be performed in a dedicated follow-up project (Phosrisom et al. 2023, in prep.).

\begin{figure}
  \centering
  	\includegraphics[width=\columnwidth]{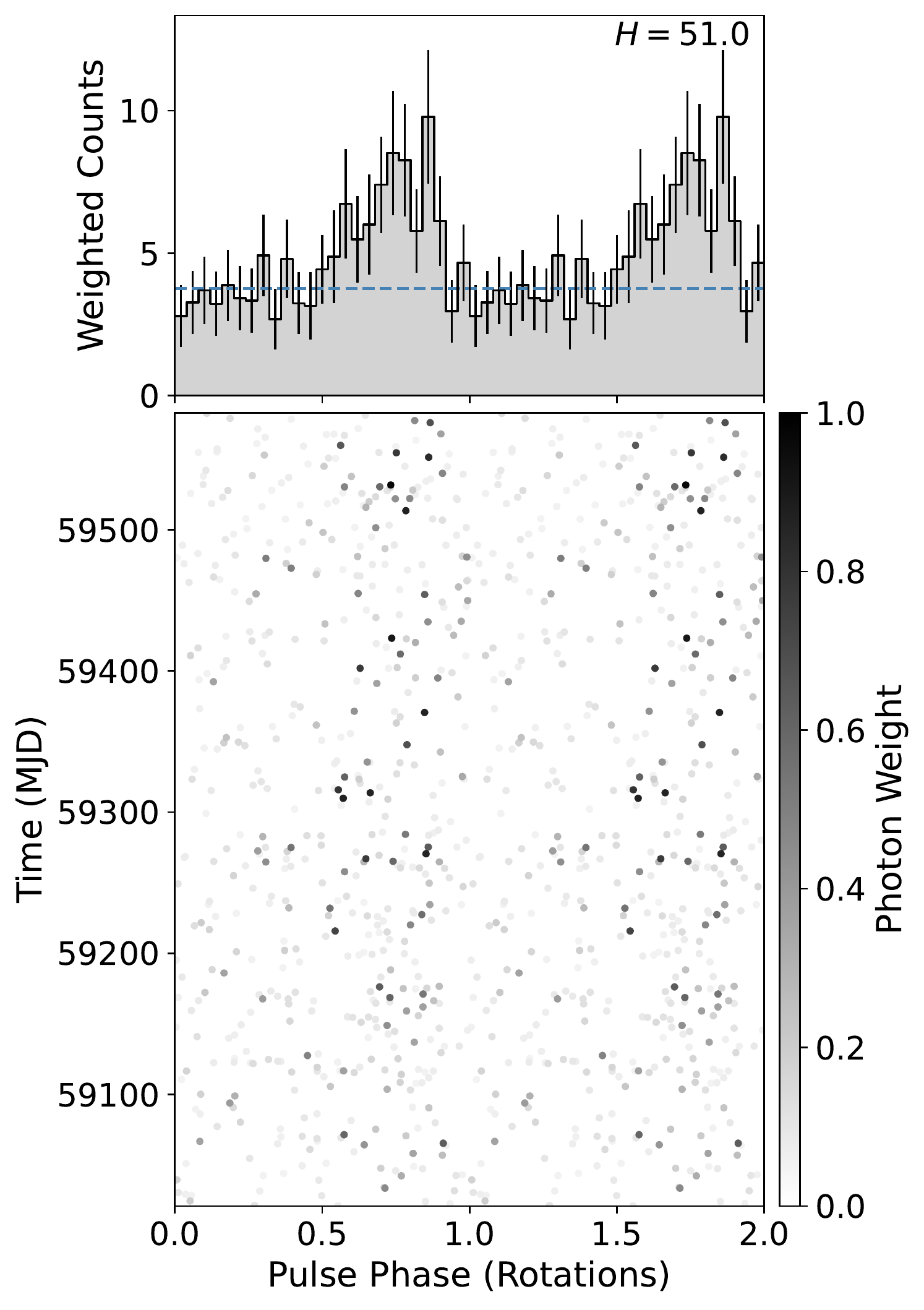}
        \caption{Gamma-ray pulsations from PSR~J1803$-$6707 in \textit{Fermi}-LAT data. The lower panel shows each photon arrival during the radio timing solution's validity interval. Each photon is assigned a weight, computed from its incidence angle and energy, and illustrated by greyscale representing the probability that it was emitted by the targeted source as opposed to a fore- or background source. The upper panel shows the integrated pulse profile, computed by summing the photon probability weights in each rotational phase bin. The dashed blue horizontal line shows the estimated background rate.}
    \label{f:J1803_fermi}
\end{figure}

As with PSR~J1526$-$2744, we folded the gamma-ray data using the radio ephemeris to check for gamma-ray pulsations. Significant pulsations are detected within the 1\,yr interval in which the radio ephemeris is valid, with a weighted $H$-test of $H=51$, corresponding to a $5.6\sigma$ detection using the false-alarm probability calibration from \citet{Bruel2019+Weights}. These pulsations are shown in Figure~\ref{f:J1803_fermi}. However, these pulsations quickly disappear when extrapolating to earlier data. This is not unexpected, as Taylor-series models for orbital period variations lack predictive power, and the spin-down rate is not measured precisely enough in the radio data to fold the 13\,yr of \textit{Fermi}-LAT data. We have not been successful in extending the ephemeris by timing the gamma-ray data, likely as a result of the pulsar's faint gamma-ray flux and the large amplitude of the orbital phase variations. A longer radio timing baseline will hopefully solve this problem by providing a more precise spin-down rate measurement, and a longer phase-connected radio ephemeris that will fold more \textit{Fermi}-LAT data to provide better statistics with which to build a gamma-ray pulse profile template to search with. Nevertheless, this detection confirms the association between 4FGL~J1803.1$-$6708 and PSR~J1803$-$6707. The gamma-ray energy flux from this source above 100\,MeV is $G_{100} = 5.0\pm 0.5\times10^{-12}$\,erg\,cm$^{-2}$\,s$^{-1}$, corresponding to an efficiency $\eta  = 1.7\%$ for the assumed \citetalias{YMW16} distance of 1.4\,kpc (for $\textrm{DM}=38.382$\,pc\,cm$^{-3}$), which is again fairly typical of gamma-ray MSPs \citep{2PC}.

\begin{figure}
 \centering
\includegraphics[width=1.0\linewidth]{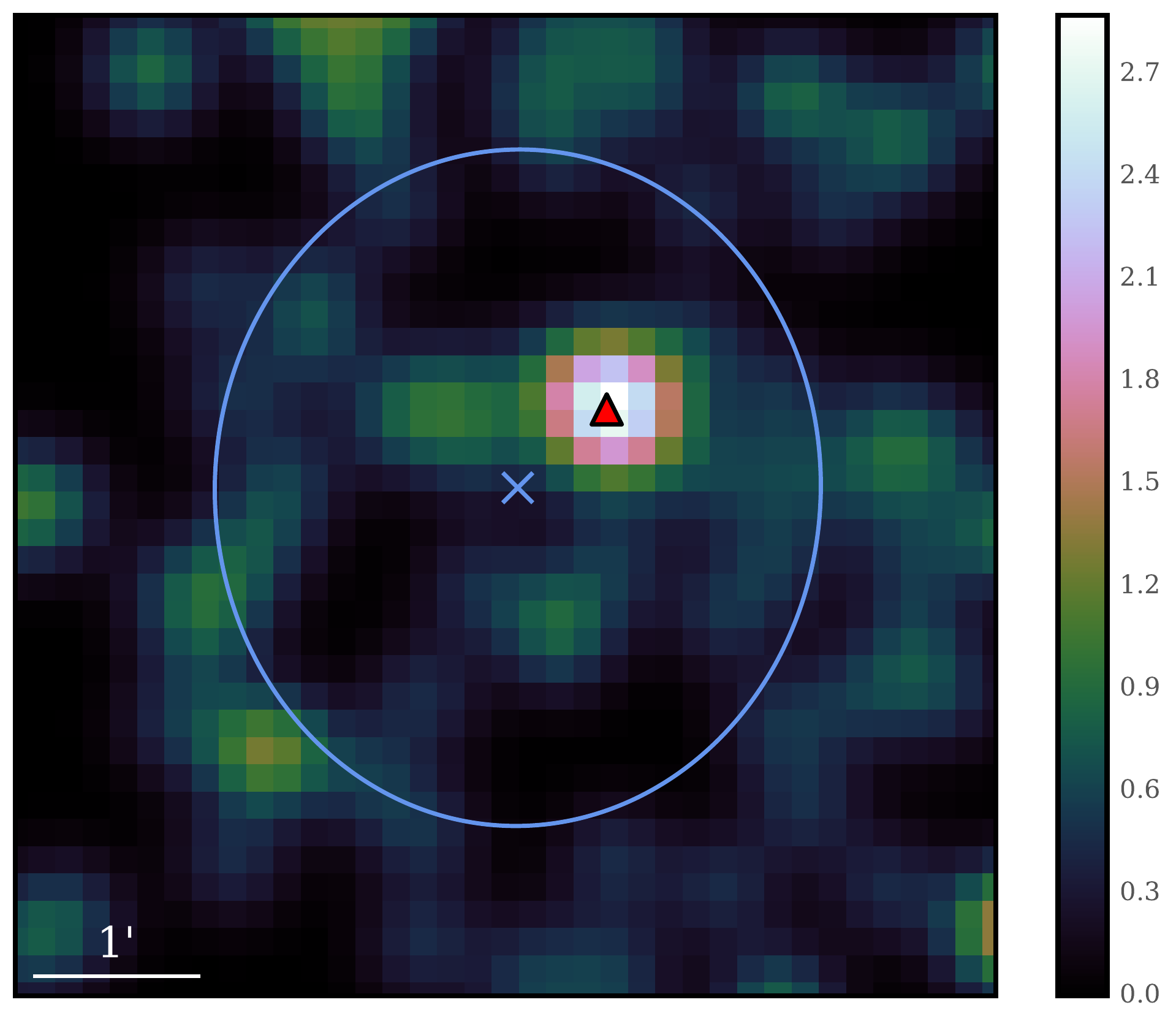}
\caption{eRASS1 image in the $0.2-2.3\,\mathrm{keV}$ band, with the $95\%$ positional error ellipse of 4FGL J1803.1$-$6708 indicated in blue, and the GAIA position overlaid on the eROSITA counterpart of PSR J1803$-$6707 marked in red. The image has been spatially binned to a pixel size of $10^{\prime\prime}$ and smoothed with a $15^{\prime\prime}$ kernel. The color bar indicates the count rate in units of $10^{-3}\,\rm{ct\,s^{-1}\,pix^{-1}}$.}
\label{f:eROSITA_Image}
\end{figure}

Using data from the first \emph{eROSITA} all-sky survey \citep[eRASS1,][]{Predehl21}, the X-ray counterpart of PSR J1803$-$6707 was independently detected (positional match within $\sim$ 5$^{\prime\prime}$) in a pilot search
for likely X-ray counterparts of unassociated \emph{Fermi-LAT} sources. Figure~\ref{f:eROSITA_Image} depicts the
X-ray counterpart as seen in eRASS1. The formal detection significance in the  $0.2-2.3\,\mathrm{keV}$ bandwidth is $4.6\sigma$ for a vignetting corrected exposure time of 167.63s and a count rate of $(7.2\pm2.4)\times10^{-2}\,\mathrm{ct\,s^{-1}}$ obtained from all seven telescope modules. The source is not 
detected in the $2.3-8.0\,\mathrm{keV}$ bandpass. The detected X-ray photons do not support a detailed spectral analysis. \textit{eROSITA}'s temporal resolution is 50 ms which prevents the detection of a periodicity at the millisecond level as well. Assuming a power-law spectrum with a photon index of 2.0 and an absorption column density of $6\times10^{20}\,\rm{cm^{-2}}$ \citep{HI4PI}, the \textit{eROSITA} counting rate implies an unabsorbed X-ray flux of $F_{\rm X} = (1.5\pm0.5)\times10^{-13}\,\mathrm{erg\,s^{-1}\,cm^{-2}}$ in the $0.2-10\,\mathrm{keV}$ band. 
For the \citealt{YMW16} distance we obtain an X-ray luminosity of $L_{\rm X} \sim 3.5\times10^{31}\,\mathrm{erg\,s^{-1}}$ and an X-ray efficiency of $L_{\rm X}/\dot{E} \sim 4.7\times 10^{-4}$, a typical value for X-ray detected MSPs \citep[see e.g.][]{Becker97}. 
A significant contribution by the $g\sim20$ optical counterpart to the detected X-ray flux is not expected. The X-ray-to-optical flux ratio of the system is around 1, much greater than that expected for coronal X-ray emission from stars \citep{Krautter99}. 

It is interesting to note that, considering the $\gamma$-ray flux of the likely counterpart, the $\gamma$-ray to X-ray flux ratio of this source is comparatively small at $F_{\gamma}/F_{\rm X} \sim 35$. This value is located at the lower end of the observed distribution for high-energy pulsars, which extends over a range of around $10-10^4$ \citep{Marelli11,Berteaud2021+XrayMSPs}. It is therefore likely that the X-ray flux contains a contribution from, or is dominated by, an intra-binary shock, as is commonly seen in redback binary systems \citep[e.g.][]{Roberts2015+RBXrays}. Follow-up observations are planned to further investigate the X-ray emission from this system.

\begin{table}
  \centering
  \caption{Radio timing solution for PSR~J1803$-$6707. Parameter values are in the TDB scale.}
  \label{t:J1803}
  \begin{tabular}{lc}
    \hline
    Parameter & Value \\
    \hline
    \multicolumn{2}{c}{\textit{Gaia} astrometry}\\
    \hline
    R.A., $\alpha$ (J2000) & $18^h03^m04\fs235339(9)$ \\
    Decl., $\delta$ (J2000) & $-67\degr07\arcmin36\farcs1576(2)$\\
    Proper motion in R.A., $\mu_{\alpha}\cos{\delta}$ (mas yr$^{-1}$) & -8.4(2)\\
    Proper motion in Decl. $\mu_{\delta}$ (mas yr$^{-1}$) & -6.5(2)\\
    Parallax, $\varpi$ (mas) & 0.18(17)\\
    Position reference epoch (MJD) & 57388.0\\
    \hline
    \multicolumn{2}{c}{Timing parameters}\\
    \hline
    Data span (MJD) & $59020$--$59583$\\
    Reference epoch (MJD) & 59364.893677\\
    Dispersion measure, DM (pc cm$^{-3}$) & 38.382(3)\\
    Spin frequency, $\nu$ (Hz) & $468.46771214886(7)$\\
    Spin-down rate, $\dot{\nu}$ (Hz s$^{-1}$) & $-4.01(2)\times10^{-15}$\\
    Orbital period, $P_{\rm orb}$ (d) & $0.38047324(8)$\\
    Projected semi-major axis, $x$ (lt s) & $1.061910(3)$\\
    Epoch of ascending node, $T_{\rm asc}$ (MJD) & $59020.99710(1)$\\
    1st orbital frequency derivative, $f_{\rm orb}^{(1)}$ (Hz s$^{-1}$) & $1.05(9)\times10^{-19}$\\
    2nd orbital frequency derivative, $f_{\rm orb}^{(2)}$ (Hz s$^{-1}$) & $-1.5(1)\times10^{-24}$\\
    3rd orbital frequency derivative, $f_{\rm orb}^{(3)}$ (Hz s$^{-1}$) & $1.03(7)\times10^{-31}$\\
    4th orbital frequency derivative, $f_{\rm orb}^{(4)}$ (Hz s$^{-1}$) & $-3.3(2)\times10^{-40}$\\
    \hline
    \multicolumn{2}{c}{Derived parameters}\\
    \hline
    Spin period, $p$ (ms) & $2.1346188308539(3)$\\
    Spin period derivative, $\dot{p}$ & $1.828(7)\times10^{-20}$\\
    Spin-down power, $\dot{E}$ (erg/s) & $7.4\times10^{34}$\\
    Surface magnetic field strength, $B_{\rm S}$ (G) & $2\times10^8$\\
    Light-cylinder magnetic field strength, $B_{\rm LC}$ (G) & $1.9\times10^5$\\
    Minimum companion mass, $M_{\rm min}$ ($M_{\odot}$) & 0.26\\
    Distance (from \citetalias{YMW16}), $d$ (kpc) & 1.4\\
    \hline
  \end{tabular}
\end{table}

\subsubsection{PSR J1036$-$4353}
This pulsar is both the fastest-spinning (with $\nu \approx 595$~Hz), and the most highly-accelerated (with $a \approx 15$\,m s$^{-2}$) pulsar found by our survey. Follow-up observations have revealed this to also be a redback system. A preliminary orbital solution revealed a $\sim$6.2\,hr orbit and a minimum companion mass of $0.23$\,M$_{\odot}$. Several non-detections have occurred during our timing campaign in the half of the orbit in which the pulsar is behind its companion, likely due to eclipses by material from the redback companion. A single \textit{Gaia} source lies within its \texttt{SeeKAT} localisation region. Like PSR~J1803$-$6707, we observed this source with ULTRACAM, revealing a double-peaked light curve, shown in Figure~\ref{f:ultracam}, very typical of redback binary systems. Modelling of this data will also be presented in a future paper (Phosrisom et al. 2023, in prep.). 

\subsubsection{The other new MSPs}
Of the remaining MSPs, one (PSR~J1709$-$0333) appears to be isolated, while three (PSRs~J1623$-$6939, J1823$-$3543 and J1858$-$5422) have companion masses and orbital periods that are consistent with helium white dwarf (He-WD) companions, but the remaining two warrant further mention. The first, PSR J1906$-$1754 lies at an unusual position in the orbital parameter space shown in Figure~\ref{f:mass_vs_PB}, with a minimum companion mass ($M_{\rm c} > 0.05\,M_{\odot}$) much lower than that of typical white-dwarf companions, but with an orbital period ($P_{\rm orb} = 6.5\,\textrm{d}$) much longer than a typical black-widow system. Of the MSP binaries with $P_{\rm orb} > 1\,\textrm{d}$, only PSR~J1737$-$0811 has a similarly low minimum companion mass \citep[$M_{\rm c} > 0.06\,M_{\odot}$,][]{Boyles2013+GBTdrift}. The low minimum mass could indicate a very face-on binary inclination angle, with $i < 14.5\deg$ required for $M_{\rm c} = 0.2\,M_{\odot}$. However, such orbits are \textit{a priori} unlikely, as only around 3\% of orbits should have a lower inclination, assuming a random distribution of viewing angles. The final pulsar, PSR~J1757$-$6032, has a larger minimum companion mass ($M_{\rm c} > 0.45$\,M$_{\odot}$) that suggests it is perhaps more likely to have a CO-WD companion, similar to that of the relativistic binary PSR~J1614$-$2230 \citep{Demorest2010+J1614,Tauris2011+J1614}. 

\subsection{Sensitivity}
\label{s:sensitivity}
As our observations were not flux-calibrated, we estimated the flux density thresholds ($S$) above which a pulsar should have been detectable to our survey. To estimate this, we used the pulsar radiometer equation \citep{PSRHandbook},
\begin{equation}
  S = \frac{\rho \, (T_{\rm sys} + T_{\rm sky})}{\beta \, G\, n_{\rm ant} \, \, \sqrt{n_{\rm pol} T_{\rm obs} B}} \sqrt{\frac{w}{1 - w}}\,
  \label{e:radiometer}
\end{equation}
where we have assumed: $G=0.044$~K~Jy$^{-1}$ is the gain per antenna \citep{Bailes2020+MeerTIME}; $\rho=9$ was the S/N threshold required for a candidate to be folded for visual inspection; $n_{\rm ant}$ is the number of antennas; $n_{\rm pol}=2$ polarisations are summed; $T_{\rm obs}$ is the exposure time; $B=700\,$MHz is the estimated useable bandwidth after RFI excision; and $T_{\rm sys} = 22.5$\,K is the combination of the receiver temperature, atmospheric and ground spillover contributions \citep{Ridolf2021+TRAPUMGCs}. The sky temperature, $T_{\rm sky}$, towards each source is taken from the 408~MHz all-sky map of \citet{Haslam1982}, reprocessed by \citet{Remazeilles2015}, and scaled to the central frequency of 1284~MHz according to an assumed spectral index of $-2.6$. The resulting flux-density limits are given in Table~\ref{t:targets}. Estimated sensitivity also depends strongly on the assumed pulse duty cycle, $w$. For the new MSPs, this varies from $5\%$ (PSR~J1803$-$6707) to $35\%$ (PSR~J1906$-$1754); we have assumed $w=15\%$ for our estimated sensitivities.

The remaining factor in Equation~\ref{e:radiometer}, $\beta$, accounts for various losses incurred during observing and searching. This includes constant fractional losses due to the 8-bit voltage digitisation (5\%) and beamforming efficiency (5\%); sensitivity losses due to using incoherent harmonic summing (15\% for our assumed 15\% duty cycle, \citealt{Morello2020+FFA}); finite time resolution and intra-channel dispersion smearing (both <1\%). In addition to these fractional losses, there are random losses due to the location of a signal between FFT bins (averaging 8\%), DM trials and acceleration trials. We estimate these losses for an assumed $\nu=500$~Hz, $\textrm{DM}=100$~pc~cm$^{-3}$ pulsar, via a Monte Carlo procedure - generating signals with random offsets from the nearest search trial and evaluating the S/N losses. The final loss factor to consider is that due to the angular offset between a pulsar and the centre of the nearest coherent beam. To evaluate this, for each source we drew random locations from the \citetalias{4FGL} localisation probability densities and evaluated the sensitivity at that location of the nearest coherent beam according to the simulated PSF model from \texttt{Mosaic}. Combining all of these losses provided a Monte Carlo distribution for $\beta$. 

For each source we quote three estimated sensitivities: $S_{\rm opt}$ is the optimum sensitivity, assuming only the constant losses described above, and therefore estimates a fundamental flux density limit for our survey; while $S_{\rm 50}$ and $S_{\rm 95}$ are the flux density limits obtained using Equation~\ref{e:radiometer} using the median and 95th centile values, respectively, of the $\beta$ values obtained from our Monte Carlo estimates. Average sensitivities were $S_{\rm opt} \approx 55\upmu$Jy, $S_{\rm 50} = 120\,(90)\,\upmu$Jy, and $S_{\rm 95} = 150\,(110)\,\upmu$Jy for the first (second) pass, respectively. The flux density threshold for the incoherent beam (for pulsars lying outside the region tiled by coherent beams) is approximately $\sqrt{n_{\rm ant}} \approx 8$ larger than that for a coherent beam. 

\section{Discussion}
\label{s:discussion}
In this paper, we have presented the first MeerKAT survey for new radio pulsars in unassociated \textit{Fermi}-LAT sources. While our strategy of targeting pulsar-like gamma-ray sources is certainly not novel, the capabilities of this next-generation radio telescope do lead to significant advantages over previous surveys.

Foremost among these is the extremely high sensitivity of the full MeerKAT array, surpassed only by the Arecibo \citep{Cromartie2016+Arecibo} and FAST \citep{Wang2021+J0318} telescopes, but unprecedented in the Southern Hemisphere. In Section~\ref{s:sensitivity}, we estimated typical 95\% flux density upper limits of $\sim$100\,$\upmu$Jy, which can be compared to the nominal $\sim$200\,$\upmu$Jy sensitivity (not including many of the loss terms that we consider) that was achieved with typical hour-long pointings in the similar survey of \textit{Fermi}-LAT sources performed at Parkes by \citet{Camilo2015+Parkes}. 

Of the newly discovered pulsars, only PSR~J1803$-$6707 is bright enough that it could perhaps have been discovered in hour-long Parkes observations, but such a discovery would have been complicated by jerk effects from its short orbital period and high acceleration. All of the new pulsars have eventually been detected in dedicated Parkes observations, but with low S/Ns that would be hard to detect without prior knowledge of the DM and spin period. 

A key factor for our survey, as \citet{Cromartie2016+Arecibo} also discussed for their survey of \textit{Fermi}-LAT sources performed with the Arecibo telescope, is that these competitive flux-density thresholds are achieved with only short 10-min observations, bringing several benefits. First, these short observation times allow for more sources to be observed within a given observing time budget, while MeerKAT's rapid ($\sim$30\,s) slew time ensures that costly overheads are not incurred by doing so. Second, short observations enable the search processing to be performed quickly (the computing cost of an acceleration search scales with at least $T^3 \log(T)$ for integration time $T$), allowing storage space to be freed quickly enough that observation scheduling is not limited by this factor, and allowing us to search up to high accelerations. Third, short observation times enable the detection of very short-period binary MSPs. For very short orbits, the assumptions that go into an acceleration search (that the orbital motion within the observation can be approximated with a constant acceleration) break down as higher-order ``jerk'' terms become significant, with sensitivity to binary MSPs only maintained for observations lasting less than 10\% of an orbit \citep{JohnstonKulkarni1991+accel}. Our short 10\,min observation strategy mitigates the worst of the jerk effects, but still means that sensitivity is lost for binaries with periods shorter than $\sim$100\,min. Only three Galactic field MSPs with shorter orbital periods than this are known, although all three are gamma-ray MSPs \citep{Pletsch2012+J1311,Stovall2014+GBNCC,Nieder2020+J1653}.

Additionally, short observations mean that multiple passes can be performed to minimise missed discoveries due to the time-varying effects that can contribute to a pulsar's detectability in a given observation. Interstellar scintillation introduces time- and frequency-dependent variations in the observed spectrum, which can lead to a pulsar being undetectable over large frequency ranges \citep{Camilo2015+Parkes}. Indeed PSR~J1709$-$0333 remained undetected in our L-band confirmation observation, despite a dense beam tiling that covered its now-known position. Our UHF confirmation observations reveal the reason for this - in both observations, separated by 70\,min, the pulsar was only detected in $\sim$20$\%$ ($\sim$100\,MHz) of the bandwidth, indicative of scintillation. Subsequent detections with the Parkes UWL and the Effelsberg L-band receivers revealed similar behaviour. 

MSPs in black-widow and redback binaries can also be undetectable for large fractions of an orbit, usually (but not exclusively) around the pulsar's superior conjunction, as a result of diffuse intra-binary plasma dispersing, scattering and absorbing radio pulsations. The jerk terms that limit the sensitivity of an acceleration search to short period binaries also have an orbital phase dependence \citep{JohnstonKulkarni1991+accel}, and so sensitivity also depends on the orbital phase at which observations take place. Our short exposure, two-pass survey partially mitigates these effects. The case of PSR~J1526$-$2744 further illustrates this: we were unable to detect this pulsar in dedicated follow-up observations with Parkes and Nan\c{c}ay prior to obtaining an orbital solution without performing jerk searches, a computationally expensive technique \citep{Andersen2018+Jerk} that has only recently started being employed in radio surveys of \textit{Fermi} sources.

One additional time-dependent effect that is specific to this survey is the fact that the shape of a coherent tied-array beam on the sky depends on the elevation at which a source is observed, with beams being more elongated for sources at lower elevations due to the smaller projected baselines between antennas. The beam tiling pattern used to cover a given source region therefore depends on the exact sidereal time and array configuration. 

Indeed, four of the nine pulsars discovered here were only detected in the second observation of their respective \textit{Fermi} sources. This was partially due to the improved flexibility of FBFUSE's tiling patterns that was developed between the first and second passes, and our use of a coarser frequency resolution but larger number of coherent beams, which allowed us to cover a larger solid angle around each source with a more sensitive tiling pattern. Two of these pulsars, PSRs~J1709$-$0333 and J1036$-$4353, lay just outside the corresponding \textit{Fermi}-LAT source 95\% localisation regions that we aimed to tile with coherent beams in the first pass, but were detected in the second pass where we targeted a larger nominal 99\% confidence region. The other two pulsars that were only detected in the second pass, PSRs~J1623$-$6936 and J1757$-$6032, lay in a less sensitive location in the coherent beam tiling pattern in their first observation, midway between three neighbouring beams, where sensitivity was $\sim$50\% of that at the centre of a beam, but were at a more favourable location in the second observation.

One final advantage to our survey, over projects using single-dish telescopes, is the rapid and precise localisation that can be obtained from a multi-beam detection using \texttt{SeeKAT}. This has two significant scientific benefits: rapid localisations enable immediate multi-wavelength follow-up and catalogue searches; and precise knowledge of a pulsar's location greatly assists in obtaining a phase-connected timing solution to fully exploit the scientific potential of a new MSP discovery. In the absence of an interferometric localisation, it often takes lengthy timing campaigns to reveal the location of a pulsar, as astrometric parameters can be highly degenerate with a pulsar's spin-down rate (and sometimes orbital parameters) until a data set spanning several months has been obtained. The \texttt{SeeKAT} positions for PSRs J1803$-$6707 and J1036$-$4353, were precise enough to unambiguously link these pulsars to their \textit{Gaia} counterparts, establishing these as redback binary systems, and providing sub-milliarcsecond astrometric uncertainties that are sufficiently precise that these parameters no longer needed to be fit for when building a timing solution. For PSR~J1526$-$2744, the $\sim$2$\arcsec$ positional uncertainty represents a reduction of a factor of $\sim$10,000 in the number of sky positions that had to be searched to detect gamma-ray pulsations, greatly decreasing the required computational effort.

With only two pointings towards our targeted sources there is still a high chance that some otherwise detectable pulsars among our targets may have been missed due to scintillation or eclipsing. We have therefore planned another two observations towards the sources observed here using the UHF receiver. All 9 pulsars discovered here were also detected at a higher S/N in our UHF observations, illustrating the sensitivity gains that can be made by observing at lower frequencies to exploit typical MSP spectra that decrease with frequency \citep{Frail2016+GMRTSpectra, Jankowski2018+Indices}, and MeerKAT's extremely low RFI environment at UHF \citep{Bailes2020+MeerTIME}. 

It is important to consider what prospects remain for further detections among \textit{Fermi}-LAT sources. To obtain a crude estimate for the number of pulsars remaining in this sample, we can simply sum the probabilities of each source being a gamma-ray pulsar that the Random Forest classifier computed in Section~\ref{s:targets}, obtaining $\sum P(\textrm{psr}) = 38.4$. Subtracting from this sum the 9 new pulsars discovered here, plus the 3 others mentioned in Section~\ref{s:results} that were recently discovered by other projects, gives $\sim$26 pulsars remaining in this sample, and around 190 pulsars in the full \citetalias{4FGL} catalogue. However, this is certainly an overestimate, as many of these sources have been searched already with other telescopes, and therefore our target list represents the remainder from a larger, unbiased sample from which more easily detected pulsars have already been removed. There are hints of this effect in our results: five of the nine new pulsars had fairly low pulsar probabilities $P(\rm psr) < 25\%$, perhaps because the more promising sources have already been surveyed extensively by other telescopes, while the lower probability targets tend to be fainter, newer sources many of which have not been searched before. This also suggests that further searches of a larger number of apparently less promising \textit{Fermi}-LAT sources below the $P(\rm psr)$ threshold that we used here may yet bear fruit. 

Furthermore, even if this is an accurate estimate of the number of gamma-ray pulsars in our sample, it does not mean that there are 26 detectable radio pulsars in this sample. A large fraction of the young gamma-ray pulsars detected by the \textit{Fermi}-LAT remain undetected in radio observations (and indeed the two recently-discovered young gamma-ray pulsars that were covered in our survey were not detected), presumably due to the viewing angle missing the radio beam. While we deliberately avoided sources at low Galactic latitudes in order to reduce the chance of young pulsars entering our sample, we note that there are 17 sources in our list at lower Galactic latitudes than 4FGL~J0802.1$-$5612, one of the two recently discovered young pulsars in our sample. Although MSPs tend to have wider radio beams, and hence are less likely to be ``radio-quiet'', a handful of gamma-ray MSPs have now been discovered in the \textit{Fermi}-LAT data, but have remained undetected in deep radio follow-up searches \citep{Clark2018+EAHMSPs,Nieder2020+J1653}, and similar objects may exist within our targets.

Nevertheless, it is certain that there are still several detectable MSPs lurking in our target list, and iterative \citetalias{4FGL} releases \citep{Ballet2020+4FGLDR2, 4FGL-DR3} bring new, albeit fainter, unassociated sources to target. Indeed, since our survey began, likely MSP binaries have been discovered in optical and X-ray searches of three of our target sources, 4FGL~J0540.0$-$7552 \citep{Strader2021+J0540}, 4FGL J0940.3$-$7610 \citep{Swihart2021+J0940}, and 4FGL~J1120.0$-$2204 \citep{Swihart2022+J1120}, while a fourth, 4FGL~J1702.7$-$5655, has recently been identified through the detection of gamma-ray eclipses \citep{Corbet2022+J1702}. 4FGL~J1120.0$-$2204 in particular appears to have a white dwarf companion in a fairly long (15\,hr) orbit, but has remained undetected in many previous long radio observations, as well as in our 10-minute pointings. Its non-detection is therefore unlikely to be explained by orbital or eclipsing effects, suggesting that deeper searches, in addition to our continuing ``shallow'' survey, may be necessary to uncover the nature of many of the remaining unidentified but pulsar-like \textit{Fermi}-LAT sources. 

\section{Conclusions and Future Work}
\label{s:conclusions}
We have presented the first results from TRAPUM's survey for new pulsars in unassociated \textit{Fermi}-LAT sources using the MeerKAT radio telescope, discovering nine new MSPs, of which eight are in binary systems. Our results continue the trend of short-period binaries being discovered at a far higher rate within \textit{Fermi}-LAT sources than in untargeted surveys, with the discovery of two new redback binaries (PSRs~J1036$-$4353 and J1803$-$6707) with optically bright companion stars and radio eclipses, and a third (PSR~J1526$-$2744) that is possibly the most compact known MSP-WD binary system in the Galactic field. Two other MSPs have preliminary orbital solutions that mark their companions as outliers among the known MSP-binary population, PSR~J1757$-$6032 appears to have a less typical CO-WD companion, while PSR~J1906$-$1754 has a much longer orbital period than any known black widow, but a minimum companion mass that is far lighter than that of a typical WD companion.

We obtained phase-connected timing solutions for two of the new MSPs, PSRs~J1526$-$2744 and J1803$-$6707, using timing observations at Parkes and Nan\c{c}ay, that enabled us to detect gamma-ray pulsations in the \textit{Fermi}-LAT data. Orbital period variations in PSR~J1803$-$6707 precluded further extrapolation of the timing solution, but for PSR~J1526$-$2744, gamma-ray timing provides a full 14-yr timing solution. This in turn enabled us to search for continuous gravitational waves from this pulsar in the Advanced LIGO O1, O2 and O3 data, although none were detected and the strain upper limit remains well above the pulsar's spin-down luminosity budget. A dedicated timing campaign is underway at MeerKAT to obtain phase-connected timing solutions, and afterwards gamma-ray pulsations, from the remaining seven discoveries (Burgay et al., 2023, in prep.), as are optical follow-up observations and modelling of the new redback binaries (Phosrisom et al., 2023, in prep.). 

Our results also emphasise the promise of continued radio surveys of \textit{Fermi}-LAT sources: several high-confidence pulsar candidates still remain within our target list but have eluded detection in ours and earlier surveys, perhaps due to scintillation, eclipses or simple intrinsic faintness. Indeed, our TRAPUM survey continues to make discoveries beyond the first results presented here. Two further passes of the sources surveyed here are currently underway using MeerKAT's UHF receiver, from which we have discovered 8 additional MSPs already\footnote{\url{http://trapum.org/discoveries/}}. We have also performed several hour-long observations of a group of high confidence redback candidates that were previously identified in optical observations, detecting a further three new MSPs that will be presented in a dedicated paper (Thongmeearkom et al., 2023, in prep.). We will continue our initial survey strategy of observing a large number of sources with short (10-minute) observations, expanding our target list to include more sources from the most recent \citetalias{4FGL-DR3} catalogue, but deeper observations may be required to fully explore the most promising gamma-ray sources.  

\section*{Acknowledgements}
The MeerKAT telescope is operated by the South African Radio Astronomy Observatory, which is a facility of the National Research Foundation, an agency of the Department of Science and Innovation. We thank staff at SARAO for their help with observations and commissioning. TRAPUM observations used the FBFUSE and APSUSE computing clusters for data acquisition, storage and analysis. These clusters were funded and installed by the Max-Planck-Institut f\"{u}r Radioastronomie (MPIfR) and the Max-Planck-Gesellschaft. The National Radio Astronomy Observatory is a facility of the National Science Foundation operated under cooperative agreement by Associated Universities, Inc. The Parkes radio telescope is part of the Australia Telescope National Facility (https://ror.org/05qajvd42) which is funded by the Australian Government for operation as a National Facility managed by CSIRO. We acknowledge the Wiradjuri people as the traditional owners of the Observatory site. The Nan\c cay Radio Observatory is operated by the Paris Observatory, associated with the French Centre National de la Recherche Scientifique (CNRS) and Universit\'{e} d'Orl\'{e}ans. It is partially supported by the Region Centre Val de Loire in France. Partly based on observations with the 100-m telescope of the MPIfR at Effelsberg.

The Fermi LAT Collaboration acknowledges generous ongoing support from a number of agencies and institutes that have supported both the development and the operation of the  LAT  as  well  as  scientific  data  analysis.  These  include  the National  Aeronautics  and Space   Administration   and   the   Department   of   Energy   in the   United   States,   the Commissariat \`{a} l'Energie Atomique and the Centre National de la Recherche Scientifique /  Institut  National  de  Physique  Nucl\'{e}aire et  de  Physique  des  Particules  in  France,  the Agenzia  Spaziale  Italiana and the Istituto  Nazionale  di  Fisica  Nucleare  in  Italy,  the Ministry  of  Education,  Culture, Sports,  Science  and  Technology  (MEXT),  High  Energy Accelerator  Research Organization  (KEK)  and  Japan  Aerospace  Exploration  Agency (JAXA)  in  Japan, and  the  K.  A.  Wallenberg  Foundation,  the  Swedish  Research  Council and the Swedish National Space Board in Sweden. 

Additional   support   for   science   analysis   during   the   operations phase is gratefully acknowledged from the Istituto Nazionale di Astrofisica in Italy and the Centre National d'Etudes Spatiales  in  France. This  work  performed  in  part  under  DOE  Contract  DE-AC02-76SF00515.

eROSITA is the primary instrument aboard SRG, a joint Russian-German science mission supported by the
Russian Space Agency (Roskosmos), in the interests of the Russian Academy of Sciences 
represented by its Space Research Institute (IKI), and the Deutsches Zentrum für Luft- und Raumfahrt
(DLR). The SRG spacecraft was built by Lavochkin Association (NPOL) and its subcontractors, and is
operated by NPOL with support from IKI and the Max Planck Institute for Extraterrestrial Physics (MPE).
The development and construction of the eROSITA X-ray instrument was led by MPE, with contributions from
the Dr.~Karl Remeis Observatory Bamberg \& ECAP (FAU Erlangen-N\"urnberg), the University of Hamburg 
Observatory, the Leibniz Institute for Astrophysics Potsdam (AIP), and the Institute for Astronomy and
Astrophysics of the University of T\"ubingen, with the support of DLR and the Max Planck Society. 
The Argelander Institute for Astronomy of the University of Bonn and the Ludwig Maximilians Universit\"at 
Munich also participated in the science preparation for eROSITA. The eROSITA data shown here were
processed using the \textsc{eSASS/NRTA} software system developed by the German eROSITA consortium.

This research has made use of data or software obtained from the Gravitational Wave Open Science Center (gw-openscience.org), a service of LIGO Laboratory, the LIGO Scientific Collaboration, the Virgo Collaboration, and KAGRA. LIGO Laboratory and Advanced LIGO are funded by the United States National Science Foundation (NSF) as well as the Science and Technology Facilities Council (STFC) of the United Kingdom, the Max-Planck-Society (MPS), and the State of Niedersachsen/Germany for support of the construction of Advanced LIGO and construction and operation of the GEO600 detector. Additional support for Advanced LIGO was provided by the Australian Research Council. Virgo is funded, through the European Gravitational Observatory (EGO), by the French Centre National de Recherche Scientifique (CNRS), the Italian Istituto Nazionale di Fisica Nucleare (INFN) and the Dutch Nikhef, with contributions by institutions from Belgium, Germany, Greece, Hungary, Ireland, Japan, Monaco, Poland, Portugal, Spain. The construction and operation of KAGRA are funded by Ministry of Education, Culture, Sports, Science and Technology (MEXT), and Japan Society for the Promotion of Science (JSPS), National Research Foundation (NRF) and Ministry of Science and ICT (MSIT) in Korea, Academia Sinica (AS) and the Ministry of Science and Technology (MoST) in Taiwan.

This work has made use of data from the European Space Agency (ESA) mission
{\it Gaia} (\url{https://www.cosmos.esa.int/gaia}), processed by the {\it Gaia}
Data Processing and Analysis Consortium (DPAC,
\url{https://www.cosmos.esa.int/web/gaia/dpac/consortium}). Funding for the DPAC
has been provided by national institutions, in particular the institutions
participating in the {\it Gaia} Multilateral Agreement.

Based on observations collected at the European Southern Observatory under ESO programmes 105.20RJ.001 and 105.20RJ.002.

C.J.C. and R.P.B. acknowledge support from the European Research Council (ERC)
under the European Union's Horizon 2020 research and innovation programme (grant
agreement No. 715051; Spiders). B.W.S. and M.C.B. acknowledge funding from the ERC under the European Union’s Horizon 2020 research and innovation programme (grant agreement No. 694745). M.B., A.P. and A.R. gratefully acknowledge financial support by the research grant ``iPeska'' (P.I. Andrea Possenti) funded under the INAF national call Prin-SKA/CTA approved with the Presidential Decree 70/2016.  E.D.B., C.J.C., D.J.C., W.C., M.K., V.V.K, L.N., P.V.P. and A.R. acknowledge continuing valuable support from the Max-Planck Society.  L.V. acknowledges financial support from the Dean’s Doctoral Scholar Award from the University of Manchester. V.S.D., T.R.M and ULTRACAM acknowledge the support of
the STFC. S.M.R. is a CIFAR Fellow and is supported by the NSF Physics Frontiers Center awards 1430284 and 2020265.

We would like to thank Matthew Kerr, Guillem Mart\'{i}-Devesa and David Smith for reviewing this paper on behalf of the \textit{Fermi}-LAT collaboration.  

\section*{Data Availability}
TRAPUM data products are available upon reasonable request to the TRAPUM collaboration. the \textit{Fermi}-LAT data are available from the \textit{Fermi} Science Support Center (\url{http://fermi.gsfc.nasa.gov/ssc}).

\bibliographystyle{mnras}
\bibliography{ms}

\bsp	% typesetting comment
\label{lastpage}
\end{document}